\documentclass[12pt]{article}

\usepackage[margin=2.5cm]{geometry}

\usepackage{graphicx}
\usepackage{amsmath,amssymb}
\usepackage{bm}
\usepackage{hyperref}
\usepackage{setspace}
\usepackage{xcolor}
\usepackage{booktabs}

\begin{document}

\title{Fundamental Limits of Quantum Sensors\\for Gravitational Wave Detection}

\author{Sergio Gaudio\\
\normalsize\itshape Department of Physics, Los Angeles City College, Los Angeles, CA 90029, USA\\
\normalsize\upshape E-mail: \texttt{gaudios@laccd.edu}}

\date{\today}

\maketitle

\noindent\footnotesize This is an author-created, un-copyedited version of an article published in Classical and Quantum Gravity. IOP Publishing Ltd is not responsible for any errors or omissions in this version of the manuscript or any version derived from it. The Version of Record is available online at \url{https://doi.org/10.1088/1361-6382/ae9449}. \normalsize\vspace{1em}

\begin{abstract}
\noindent
Recent advances in quantum sensing---optical clocks at $5.5\times 10^{-19}$ systematic uncertainty, frequency-dependent squeezing below the standard quantum limit, quantum magnetometers approaching fundamental sensitivity limits---raise a natural question: can these technologies detect gravitational waves directly, or enhance existing detectors beyond current capabilities? We show that the answer is primarily determined by the \emph{coupling mechanism} between the gravitational wave and the sensor. Starting from the tidal Hamiltonian in Fermi normal coordinates, we identify three physically distinct mechanisms by which a gravitational wave couples directly to a quantum system, and derive their transducer gains within linearized general relativity and non-relativistic quantum mechanics. Internal atomic coupling (tidal distortion of electronic wavefunctions) yields a transducer gain $G_A = 2.4\times 10^{-20}$, with vanishing first-order energy shifts for all $J=0$ clock states---a $\sim\!10^{35}$ deficit relative to laser interferometry that exceeds any projected quantum enhancement. Center-of-mass coupling (Doppler shifts from geodesic motion) reaches strain sensitivities of $\sim\!10^{-18}$, still $10^4$ above LISA requirements. Light propagation coupling (phase accumulation over macroscopic baselines) provides the enormous transducer gain that makes laser interferometry---and atom interferometry---viable. For detectors exploiting this third mechanism, we quantify how much improvement quantum sensors can provide through the detector's noise architecture: LISA's noise budget is predominantly classical, limiting combined quantum enhancement to $\mathcal{E} \approx 1.04$, while ground-based detectors in the shot-noise-dominated regime achieve $\mathcal{E} = 1.6$--$2.1$. Atom interferometers exploit the same light-propagation mechanism to target the 0.01--10~Hz band between the LISA and LIGO ranges.
\end{abstract}


\section{Introduction}
\label{sec:intro}

The direct detection of gravitational waves by the LIGO and Virgo collaborations~\cite{Abbott2016} opened the field of gravitational wave astronomy. The LIGO-Virgo-KAGRA (LVK) network has now observed approximately 340 candidate events through the completion of the fourth observing run (O4) in November 2025, with 218 confident detections cataloged in GWTC-4.0~\cite{GWTC3,GWTC4}. The space-based detector LISA, approved by ESA for launch in the 2030s~\cite{LISARedBook2024}, will extend observations to the millihertz band, targeting massive black hole mergers, extreme mass-ratio inspirals, and galactic binaries. Ground-based detectors are advancing toward third-generation instruments---the Einstein Telescope~\cite{ET2010} and Cosmic Explorer~\cite{CE2019}---with order-of-magnitude sensitivity improvements. Simultaneously, atom interferometric detectors target the 0.01--10~Hz mid-band gap between LISA and LIGO, a range not covered by either operating or approved instruments~\cite{MAGIS100,AION2020,ZAIGA2020}.

Quantum sensing has achieved extraordinary precision across multiple platforms. Optical atomic clocks now reach fractional frequency uncertainties of $5.5\times 10^{-19}$~\cite{Marshall2025}, squeezed light has been deployed in gravitational wave detectors with increasing sophistication from initial injection~\cite{Tse2019,Acernese2019} to frequency-dependent operation surpassing the standard quantum limit~\cite{Jia2024}, and quantum magnetometers have projected fundamental sensitivities of $\sim 10$~aT$\sqrt{\mathrm{cm}^3/\mathrm{Hz}}$ at geomagnetic fields~\cite{Dikopoltsev2025}. These advances naturally motivate the question: can quantum sensors detect gravitational waves directly, or meaningfully enhance existing detector sensitivity beyond what squeezed vacuum injection already provides?

We show that this question cannot be answered by characterizing the quantum sensor alone. The determining factor is the \emph{coupling mechanism}---the physical process by which a gravitational wave imprints itself on the sensor's observable. A gravitational wave can couple directly to a quantum system through three physically distinct mechanisms: (i) tidal distortion of internal structure (atomic wavefunctions, molecular bonds), (ii) center-of-mass motion along geodesics (Doppler shifts), and (iii) modification of light propagation over macroscopic baselines (phase accumulation). Each mechanism has a computable transducer gain---the conversion factor from strain $h$ to the sensor's native observable---and these gains differ by up to $10^{35}$. This hierarchy, which follows from general relativity and atomic physics, predetermines the viability of any detection scheme before the quantum sensor's performance becomes relevant. The coupling mechanism thus sets whether detection is feasible at all; where it is, the detector's noise architecture sets how much a quantum sensor can add---the value of a quantum sensor is determined by these two together, not by its intrinsic precision. Throughout the analysis we keep three kinds of statement distinct: what gravity permits (fixed by linearized general relativity and non-relativistic quantum mechanics, binding on every detector), what a given noise architecture allows, and what present technology can achieve. We keep these distinct so that a rigorous derivation and an order-of-magnitude projection are never mistaken for one another.

Three results follow.

\emph{First}, direct quantum detection of gravitational waves through internal atomic coupling faces a $\sim\!10^{35}$ transducer gain deficit relative to laser interferometry. This deficit arises from the small Mechanism~A transducer gain ($G_A \sim 10^{-20}$) and the exact vanishing of first-order energy shifts for all $J = 0$ clock states, and exceeds any projected quantum enhancement.

\emph{Second}, for detectors that exploit light propagation, the quantum improvement achievable is set by the detector's noise architecture. The fraction of total noise power that quantum technologies can address---which we denote $\beta$---bounds the maximum enhancement as $\mathcal{E}_{\mathrm{max}} = 1/\sqrt{1-\beta}$. For LISA ($\beta \sim 0.09$), even perfect quantum sensors yield $\mathcal{E} \approx 1.04$. For ground-based detectors at high frequencies ($\beta \sim 0.9$), squeezed vacuum provides $\mathcal{E} = 1.6$--$2.1$.

\emph{Third}, atom interferometric detectors exploit the same light-propagation coupling mechanism to access the 0.01--10~Hz decihertz band, which lies between the LISA and LIGO bands and is not covered by either operating ground-based detectors or the approved LISA mission. Other concepts have been proposed for this band~\cite{Mandel2018}; atom interferometers are distinguished within it by using freely falling atoms as inertial test masses---immune to suspension thermal noise---with common-mode rejection of laser phase noise, alongside an active demonstrator program. Quantum superposition and coherent manipulation are essential elements of the measurement.

The paper is organized as follows. Section~\ref{sec:coupling} derives the three coupling mechanisms and the transducer gain hierarchy, including the complete three-dimensional matrix element for the gravitational wave--atom interaction and its generalization to multi-electron clock atoms. Section~\ref{sec:direct} applies this to concrete direct-detection schemes using their published parameters. Section~\ref{sec:noise} develops the noise architecture analysis and applies it to LISA. Section~\ref{sec:atom_interf} presents atom interferometric detectors, which exploit the same light-propagation mechanism to access the mid-band gap. Section~\ref{sec:ground} contrasts these with ground-based detectors, and Section~\ref{sec:discussion} discusses implications across the spectrum, including the nanohertz band and quantum network proposals.

Throughout, we adopt the Misner--Thorne--Wheeler (MTW) sign convention for the Riemann tensor~\cite{MTW1973}, use SI units, and employ the metric signature $(-,+,+,+)$.

\section{Gravitational Wave Coupling to Quantum Systems}
\label{sec:coupling}

Within linearized gravity, a gravitational wave can couple directly to a quantum system in three distinct ways: through its internal structure, its center-of-mass motion, or the propagation of light between separated worldlines. These exhaust the direct channels; indirect mechanisms, in which an external field first converts the wave into another quantity (for instance, conversion of a gravitational wave into photons in a static magnetic field), lie outside our scope. We derive each below.

\subsection{Mechanism A: Internal tidal coupling}
\label{sec:mechanism_A}

\subsubsection{The tidal Hamiltonian from general relativity}
\label{sec:tidal_H}

In the proper detector frame (Fermi normal coordinates), a particle at position $x^i$ relative to a freely falling observer at the origin experiences the tidal gravitational field through the geodesic deviation equation. For non-relativistic particles, the resulting Hamiltonian is~\cite{MTW1973,Maggiore2007}
\begin{equation}
\hat{H}_{\mathrm{tidal}} = +\frac{1}{2}m\,R^i{}_{0j0}(t)\,\hat{x}^i\hat{x}^j,
\label{eq:H_tidal}
\end{equation}
where $R^i{}_{0j0}$ is the electric part of the Riemann tensor evaluated at the observer's worldline, $m$ is the particle mass, and $\hat{x}^i$ are position operators.

The derivation proceeds as follows. The geodesic deviation equation, $\ddot{\xi}^i = -R^i{}_{0j0}\,\xi^j$ in the MTW convention, describes the relative acceleration of two nearby geodesics. The force on a particle of mass $m$ is therefore $F^i = -m\,R^i{}_{0j0}\,x^j$. Identifying $F^i = -\partial U/\partial x^i$ gives the potential $U = +\frac{1}{2}m\,R^i{}_{0j0}\,x^i x^j$. The quantum version promotes $x^i$ to operators. Equation~(\ref{eq:H_tidal}) is valid to leading order in $|x|/\lambda_{\mathrm{gw}}$, which is the relevant regime for atomic systems where $|x| \sim a_0 = 5.29\times 10^{-11}$~m (the Bohr radius) and $\lambda_{\mathrm{gw}} = c/f_{\mathrm{gw}} = 3\times 10^{11}$~m at $f_{\mathrm{gw}} = 1$~mHz (the center of the LISA band)---a ratio of $a_0/\lambda_{\mathrm{gw}} \sim 2\times 10^{-22}$.

The Bohr radius enters because the position operators $\hat{x}^i$ act on the bound electron, whose matrix elements are determined by the spatial extent of the electronic wave function. For the hydrogen ground state, $\langle r^2 \rangle = 3a_0^2$. The tidal coupling strength is therefore set by $m_e \omega^2 a_0^2$: the gravitational wave attempts to produce a quadrupole distortion of a charge cloud spanning $\sim 10^{-11}$~m, using a field whose wavelength exceeds $10^{11}$~m. The atom is too small for the tidal field to grip.

\subsubsection{Plus-polarized gravitational wave}
\label{sec:plus_pol}

For a monochromatic plus-polarized gravitational wave propagating along $z$, the TT-gauge metric perturbation is $h_{xx}^{\mathrm{TT}} = -h_{yy}^{\mathrm{TT}} = h_0\cos\omega t$. The non-vanishing Riemann components in the proper detector frame are~\cite{Maggiore2007}
\begin{align}
R^x{}_{0x0} &= -\tfrac{1}{2}\ddot{h}_{xx} = +\tfrac{1}{2}\omega^2 h_0\cos\omega t, \label{eq:Rx}\\
R^y{}_{0y0} &= -\tfrac{1}{2}\ddot{h}_{yy} = -\tfrac{1}{2}\omega^2 h_0\cos\omega t, \label{eq:Ry}\\
R^z{}_{0z0} &= 0 \quad\text{(transverse wave)}. \label{eq:Rz}
\end{align}
One may verify the tracelessness $R^x{}_{0x0} + R^y{}_{0y0} + R^z{}_{0z0} = 0$, which is a direct consequence of the vacuum Einstein equation $R_{\mu\nu} = 0$.

Substituting Eqs.~(\ref{eq:Rx})--(\ref{eq:Rz}) into Eq.~(\ref{eq:H_tidal}) and simplifying:
\begin{equation}
\hat{H}_{\mathrm{tidal}} = \frac{1}{4}m\omega^2 h_0\cos(\omega t)\,(\hat{x}^2 - \hat{y}^2).
\label{eq:H_explicit}
\end{equation}
This is exact within linearized general relativity for a system of size $\ll \lambda_{\mathrm{gw}}$. The cross polarization ($h_\times$) gives the same coupling strength with the quadrupole pattern rotated by $45^\circ$; we compute for plus polarization without loss of generality.

\subsubsection{Spherical tensor decomposition}
\label{sec:spherical}

To exploit the angular momentum structure of atomic states, we decompose the operator $\hat{x}^2 - \hat{y}^2$ in spherical coordinates. Using $x = r\sin\theta\cos\phi$ and $y = r\sin\theta\sin\phi$:
\begin{equation}
\hat{x}^2 - \hat{y}^2 = \hat{r}^2\sin^2\!\theta\,\cos 2\phi.
\end{equation}
The spherical harmonics of degree $l = 2$ and order $m = \pm 2$ are
\begin{equation}
Y_2^{\pm 2}(\theta,\phi) = \frac{1}{4}\sqrt{\frac{15}{2\pi}}\sin^2\!\theta\,e^{\pm 2i\phi}.
\end{equation}
Therefore
\begin{equation}
Y_2^{+2} + Y_2^{-2} = \frac{1}{2}\sqrt{\frac{15}{2\pi}}\sin^2\!\theta\cos 2\phi,
\end{equation}
giving
\begin{equation}
\sin^2\!\theta\cos 2\phi = 2\sqrt{\frac{2\pi}{15}}\left(Y_2^{+2} + Y_2^{-2}\right).
\label{eq:cos2phi}
\end{equation}
Substituting into Eq.~(\ref{eq:H_explicit}):
\begin{equation}
\hat{H}_{\mathrm{tidal}} = \frac{1}{2}m\omega^2 h_0\cos(\omega t)\sqrt{\frac{2\pi}{15}}\,\hat{r}^2\!\left(Y_2^{+2} + Y_2^{-2}\right).
\label{eq:H_Y22}
\end{equation}
The tidal Hamiltonian is a rank-2 spherical tensor operator with components $q = \pm 2$ only (for plus polarization).

\subsubsection{Selection rules}
\label{sec:selection}

The operator $\hat{r}^2 Y_2^q$ with $q = \pm 2$ has definite tensor rank. The matrix element between atomic states $|n,l,m\rangle$ and $|n',l',m'\rangle$ factors as
\begin{equation}
\langle n'\!,l'\!,m'|\hat{r}^2 Y_2^q|n,l,m\rangle = \langle n'\!,l'|\hat{r}^2|n,l\rangle_{\!\mathrm{rad}} \times \langle l'\!,m'|Y_2^q|l,m\rangle_{\!\mathrm{ang}}.
\end{equation}
The angular integral is evaluated using the Wigner--Eckart theorem~\cite{Sakurai2017}:
\begin{equation}
\langle l'\!,m'|Y_2^q|l,m\rangle = (-1)^{m'}\!\sqrt{\frac{5(2l'\!+\!1)(2l\!+\!1)}{4\pi}}\begin{pmatrix}l' & 2 & l\\0 & 0 & 0\end{pmatrix}\!\begin{pmatrix}l' & 2 & l\\-m' & q & m\end{pmatrix}\!,
\label{eq:WE}
\end{equation}
where the arrays are Wigner $3j$ symbols. These impose three constraints.

\emph{Triangle inequality:} $|l - 2| \leq l' \leq l + 2$, giving $\Delta l = 0, \pm 1, \pm 2$.

\emph{Parity:} The first $3j$ symbol $\bigl(\begin{smallmatrix}l' & 2 & l\\0 & 0 & 0\end{smallmatrix}\bigr)$ vanishes unless $l' + 2 + l$ is even. Therefore $l' + l$ must be even, which eliminates $\Delta l = \pm 1$:
\begin{equation}
\Delta l = 0, \pm 2 \quad\text{only}.
\label{eq:delta_l}
\end{equation}

\emph{Magnetic quantum number:} $m' = m + q$, giving $\Delta m = \pm 2$ for plus polarization.

These are the selection rules for electric quadrupole (E2) transitions---the gravitational wave couples to atoms as a quadrupole field, not as a dipole field, in contrast to the spin-1 electromagnetic interaction which produces E1 selection rules ($\Delta l = \pm 1$, $\Delta m = 0, \pm 1$).

\subsubsection{Exact vanishing of first-order energy shift for clock states}
\label{sec:vanishing}

This result is the single most important finding for direct quantum detection schemes. For any $s$-state ($l = 0$, $m = 0$), the diagonal matrix element of $\hat{H}_{\mathrm{tidal}}$ is proportional to
\begin{equation}
\langle n,0,0|r^2 Y_2^q|n,0,0\rangle = \langle n,0|r^2|n,0\rangle \times \langle 0,0|Y_2^q|0,0\rangle.
\end{equation}
Since $Y_0^0 = 1/\sqrt{4\pi}$ is a constant, the angular integral is
\begin{equation}
\langle 0,0|Y_2^q|0,0\rangle = \frac{1}{4\pi}\int Y_2^q(\theta,\phi)\,d\Omega = 0,
\label{eq:angular_vanish}
\end{equation}
vanishing by the orthogonality of spherical harmonics ($l = 0$ and $l = 2$ are orthogonal). Therefore
\begin{equation}
\langle n,0,0|\hat{H}_{\mathrm{tidal}}|n,0,0\rangle = 0 \quad\text{(exact, all $n$)}.
\label{eq:first_order_vanish}
\end{equation}

The physical content of this result is transparent: a spherically symmetric charge distribution has zero quadrupole moment. The traceless tidal field, which stretches along $x$ and compresses along $y$, produces $\langle\hat{x}^2\rangle = \langle\hat{y}^2\rangle = \frac{1}{3}\langle\hat{r}^2\rangle$ for a spherically symmetric state, so $\langle\hat{x}^2 - \hat{y}^2\rangle = 0$ identically.

The consequence for atomic clocks requires a more general argument. Most precision optical clocks use transitions between states with total angular momentum $J = 0$ (e.g., ${}^1S_0 \to {}^3P_0$ in ${}^{27}$Al$^+$~\cite{Brewer2019,Marshall2025}, ${}^{87}$Sr, ${}^{171}$Yb~\cite{Ludlow2015}). For the lower clock state ${}^1S_0$, the vanishing follows from the $l = 0$ argument above. But the upper clock state ${}^3P_0$ has orbital angular momentum $L = 1$ coupled to electron spin $S = 1$ to form total $J = 0$. It is \emph{not} an $s$-state; the orbital wave function contains $p$-orbital components $|m_L = -1, 0, +1\rangle$ with Clebsch--Gordan coefficients $1/\sqrt{3}$ each. Nevertheless, the charge distribution is spherically symmetric: by the addition theorem for spherical harmonics, $\sum_{m=-l}^{l}|Y_l^m|^2 = (2l+1)/(4\pi)$, so the equal-weight superposition over all $m_L$ values produces an isotropic electron density even for $L = 1$.

The operator $\hat{r}^2 Y_2^q$ is a rank-2 spherical tensor that acts on spatial coordinates and commutes with spin. Under rotations generated by the \emph{total} angular momentum $\mathbf{J} = \mathbf{L} + \mathbf{S}$, it remains a rank-2 tensor, and the Wigner--Eckart theorem applied to $J$ gives~\cite{Sakurai2017}
\begin{equation}
\langle J'\!=\!0, M'\!=\!0|\hat{r}^2 Y_2^q|J\!=\!0, M\!=\!0\rangle \propto \begin{pmatrix} 0 & 2 & 0 \\ 0 & q & 0 \end{pmatrix} = 0
\end{equation}
because the triangle inequality $|0 - 0| = 0 < 2$ is violated. This holds for \emph{any} $J = 0$ state regardless of the values of $L$ and $S$.

The physical origin of this protection is the rank-2 tensor structure of the tidal coupling. Because the metric perturbation $h_{\mu\nu}$ carries two indices, the tidal Hamiltonian is a rank-2 tensor (Sec.~\ref{sec:spherical}). A rank-2 tensor can only have non-vanishing diagonal matrix elements between states with $J \geq 1$ (from the triangle inequality $|J - J| \leq 2 \leq J + J$, giving $J \geq 1$). The $J = 0$ clock states are therefore immune to first-order gravitational wave perturbation not by coincidence but by a selection rule that follows from the angular structure of the coupling alone.

The first-order differential clock shift is therefore
\begin{equation}
\delta\nu_{\mathrm{clock}}^{(1)} = \frac{\langle {}^3P_0|\hat{H}_{\mathrm{tidal}}|{}^3P_0\rangle - \langle {}^1S_0|\hat{H}_{\mathrm{tidal}}|{}^1S_0\rangle}{h_{\mathrm{Planck}}} = 0.
\label{eq:clock_vanish}
\end{equation}
This is not an approximation---it is an exact consequence of angular momentum algebra. Any observable gravitational wave effect on atomic clock transitions must enter through second-order perturbation theory, with an additional suppression factor of $\hat{H}_{\mathrm{tidal}}/\Delta E_{l=2}$.

We emphasize that Eq.~(\ref{eq:clock_vanish}) applies to all $J = 0 \to J = 0$ clock transitions in all atomic species. The vanishing is a selection rule, not a numerical accident. It cannot be circumvented by choosing a different atom, improving the interrogation time, or employing quantum entanglement.

\subsubsection{Explicit matrix element: hydrogen $1s \to 3d$}
\label{sec:matrix_element}

Since the first-order shift vanishes, the leading coupling is through off-diagonal matrix elements connecting $l = 0$ states to $l = 2$ states ($\Delta l = +2$). For hydrogen, the dominant transition is $|1,0,0\rangle \to |3,2,+2\rangle$ (the lowest $d$-state, since $n \geq l+1$ requires $n' \geq 3$). We compute this matrix element explicitly.

The hydrogen radial wave functions are~\cite{Sakurai2017}
\begin{align}
R_{10}(r) &= 2\,a_0^{-3/2}\,e^{-r/a_0}, \label{eq:R10}\\
R_{32}(r) &= \frac{4}{81\sqrt{30}}\,a_0^{-3/2}\left(\frac{r}{a_0}\right)^{\!2} e^{-r/(3a_0)}.
\label{eq:R32}
\end{align}

\emph{Radial matrix element.} The integral $\langle 3,2|r^2|1,0\rangle_{\mathrm{rad}} = \int_0^\infty R_{32}\,r^2\,R_{10}\,r^2\,dr$ reduces to a standard gamma function integral $\int_0^\infty r^6 e^{-4r/(3a_0)}\,dr = (3a_0/4)^7 \cdot 6!$, giving
\begin{equation}
\langle 3,2|r^2|1,0\rangle_{\!\mathrm{rad}} = 1.733\,a_0^2.
\label{eq:radial_result}
\end{equation}

\emph{Angular matrix element.} For the transition $|0,0\rangle \to |2,+2\rangle$ with operator $Y_2^{+2}$, the Wigner--Eckart theorem [Eq.~(\ref{eq:WE})] gives
\begin{equation}
\langle 2,+2|Y_2^{+2}|0,0\rangle = \frac{1}{\sqrt{4\pi}}.
\label{eq:angular_result}
\end{equation}

\emph{Full matrix element.} From Eq.~(\ref{eq:H_Y22}), the time-independent perturbation amplitude is $\hat{V}_0 = \frac{1}{2}m\omega^2 h_0\sqrt{2\pi/15}\,\hat{r}^2(Y_2^{+2} + Y_2^{-2})$. Only $Y_2^{+2}$ contributes to the $|1,0,0\rangle \to |3,2,+2\rangle$ transition (by $\Delta m = +2$):
\begin{align}
|\langle 3,2,+2|\hat{V}_0|1,0,0\rangle| &= \frac{1}{2}m\omega^2 h_0\sqrt{\frac{2\pi}{15}}\cdot 1.733\,a_0^2\cdot\frac{1}{\sqrt{4\pi}} \nonumber\\
&= \frac{1.733}{2\sqrt{30}}\,m\omega^2 h_0\,a_0^2,
\label{eq:full_ME}
\end{align}
where we have used $\sqrt{2\pi/15}\cdot 1/\sqrt{4\pi} = 1/\sqrt{30}$.

The geometric factor is
\begin{equation}
\mathcal{G} = \frac{1.733}{2\sqrt{30}} = 0.158.
\label{eq:geometric}
\end{equation}
This is the coefficient that dimensional analysis alone cannot provide. It arises from the product of three effects: the angular suppression $1/\sqrt{30}$ from the quadrupole geometry, the radial overlap integral $1.733$, and the factor $1/2$ from the tidal Hamiltonian prefactor.

\subsubsection{Transducer gain for internal atomic coupling}
\label{sec:alpha}

To compare the three coupling mechanisms on equal footing, we define a dimensionless transducer gain for each: the observable produced per unit strain. For Mechanism~A, this is the ratio of the tidal matrix element (at unit strain) to the quantum energy scale $\hbar\omega$\footnote{$G_A$ is a semiclassical transducer gain---the atomic phase shift per unit strain of a classical gravitational wave---not a fundamental coupling constant. The underlying vertex coupling between a single graviton and an electron is characterized by $\alpha_G = Gm_e^2/(\hbar c) \approx 1.75\times 10^{-45}$, the gravitational analogue of $\alpha_{\mathrm{EM}}$. For a classical wave with graviton occupation number $n \gg 1$, the semiclassical amplitude $h_0 \propto \sqrt{n}$ absorbs the factor $\sqrt{\alpha_G}$, yielding a transducer gain that depends on $\omega$ but not on $G$.}:
\begin{equation}
G_A \equiv \frac{|\langle f|\hat{V}_0|i\rangle|_{h_0=1}}{\hbar\omega} = \mathcal{G}\,\frac{m_e\omega\,a_0^2}{\hbar}.
\label{eq:alpha_def}
\end{equation}
Numerically, at $f_{\mathrm{gw}} = 1$~mHz ($\omega = 6.28\times 10^{-3}$~rad/s):
\begin{align}
G_A &= \mathcal{G}\,\frac{m_e\,\omega\,a_0^2}{\hbar} \nonumber\\
&= 0.158\times 1.52\times 10^{-19} \nonumber\\
&= 2.4\times 10^{-20},
\label{eq:alpha_numerical}
\end{align}
where $m_e\omega a_0^2/\hbar = 1.52\times 10^{-19}$ for $f_{\mathrm{gw}} = 1$~mHz.

The transducer gain is a \emph{derived quantity} from the complete three-dimensional calculation, not an assumption. A purely dimensional estimate $m_e\omega a_0^2/\hbar = 1.52\times 10^{-19}$ overestimates the coupling by a factor of $\sim 6$ due to the geometric suppression $\mathcal{G} \approx 1/6$.

\emph{Which mass?} The electron mass $m_e$ appears because $G_A$ pertains to the \emph{internal electronic transition}. The center-of-mass tidal displacement of the whole atom uses the nuclear mass $m_{\mathrm{atom}}$; replacing $m_e \to m_{\mathrm{atom}}$ in the dimensional estimate gives $\sim (m_{\mathrm{atom}}/m_e) \times 1.52\times 10^{-19} \sim 10^{-14}$ for ${}^{87}$Sr---roughly six orders of magnitude larger than $G_A$, but corresponding to a different observable (CM motion, not clock frequency shift). This distinction is critical for analyzing detection schemes.

\emph{Generality beyond hydrogen.} The explicit calculation above uses hydrogen wave functions, but the conclusions apply to all atomic species used in precision clocks (${}^{87}$Sr, ${}^{171}$Yb, ${}^{27}$Al$^+$). The angular suppression factor $1/\sqrt{30}$ originates from the rank-2 (quadrupole) structure of the tidal coupling and is species-independent. The radial matrix element $\langle r^2 \rangle$ for valence electrons in neutral atoms and singly charged ions remains of order $a_0^2$ due to electron screening: even for $Z = 70$ (Yb), the outer $6s$ electron has an effective Bohr radius $a_{\mathrm{eff}} \sim a_0$ because the nuclear charge is screened to $Z_{\mathrm{eff}} \sim 1$--$2$ by inner shells. Species-dependent corrections modify the geometric factor $\mathcal{G}$ by at most $\mathcal{O}(1)$ factors but do not alter the scaling $G_A \sim m_e \omega a_0^2/\hbar$; across species this places $G_A$ in the range $\sim(1$--$5)\times 10^{-20}$, leaving the $\sim\!10^{35}$ deficit unchanged. The hydrogen calculation therefore provides a reliable order-of-magnitude estimate that, if anything, slightly overestimates the coupling for heavier atoms whose clock states have more compact wave functions. Crucially, the exact vanishing of the first-order shift for $J = 0$ states [Eq.~(\ref{eq:clock_vanish})] is a consequence of angular momentum algebra alone and holds identically for all atomic species.

\subsubsection{Second-order energy shift}
\label{sec:second_order}

Since the first-order shift vanishes for $J = 0$ clock states, the leading observable effect arises at second order in perturbation theory. For a monochromatic perturbation $\hat{H}_{\mathrm{tidal}} = \hat{V}_0\cos(\omega t)$, the time-averaged (AC Stark) energy shift is
\begin{equation}
\delta E^{(2)} = \frac{1}{2}\sum_{n'\neq g}|\langle n'|\hat{V}_0|g\rangle|^2\!\left(\frac{1}{E_g - E_{n'} + \hbar\omega} + \frac{1}{E_g - E_{n'} - \hbar\omega}\right)\!.
\label{eq:AC_Stark}
\end{equation}
This formula follows from standard time-dependent perturbation theory: the perturbation virtually excites the atom from $|g\rangle$ to $|n'\rangle$ and back, accumulating a phase shift proportional to $|\hat{V}_0|^2$. The two denominators correspond to the co-rotating and counter-rotating terms of the time-dependent perturbation, respectively.

For gravitational waves, $\hbar\omega_{\mathrm{gw}} \sim 4\times 10^{-18}$~eV at 1~mHz while $|E_g - E_{n'}| \sim 10$~eV, so $\hbar\omega/\Delta E \sim 10^{-19}$. The denominators are indistinguishable and the sum simplifies to the static second-order formula:
\begin{equation}
\delta E^{(2)} \approx \sum_{n'\neq g}\frac{|\langle n'|\hat{V}_0|g\rangle|^2}{E_g - E_{n'}}.
\label{eq:second_order}
\end{equation}
Since $\hat{V}_0 \propto h_0$ [Eq.~(\ref{eq:H_Y22})], the matrix element squared scales as $h_0^2$: the coupling is \emph{quadratic} in strain at second order, not linear. The dominant term is $n' = 3$, $l' = 2$ with energy denominator $E_1 - E_3 = -12.09$~eV. Writing $|\langle 3,2,+2|\hat{V}_0|1,0,0\rangle|^2 = \mathcal{G}^2(m_e\omega^2 a_0^2)^2 h_0^2$ [from Eq.~(\ref{eq:full_ME})]:
\begin{align}
\delta E^{(2)} &\approx -\frac{(0.158)^2(m_e\omega^2 a_0^2)^2\,h_0^2}{12.09~\mathrm{eV}} \nonumber\\
&= -\frac{0.025\times(6.28\times 10^{-37}~\mathrm{eV})^2\,h_0^2}{12.09~\mathrm{eV}} \nonumber\\
&= -8.2\times 10^{-76}\,h_0^2~\mathrm{eV}.
\label{eq:second_order_numerical}
\end{align}
For a gravitational wave with $h_0 = 10^{-21}$: $\delta E^{(2)} = 8.2\times 10^{-118}$~eV, giving a fractional clock shift
\begin{equation}
\frac{\delta\nu^{(2)}}{\nu_0} \sim 10^{-117}
\end{equation}
for a clock transition energy of order 1~eV. This is approximately 99 orders of magnitude below the best achievable clock precision of $\sigma_y = 5.5\times 10^{-19}$~\cite{Marshall2025}.

This deficit cannot be closed by improving clock technology. Since $\delta E^{(2)} \propto h_0^2$, the signal grows with strain, but the derivation is valid only within linearized gravity ($h_0 \ll 1$). Even extrapolating to the boundary of the linearized regime, $h_0 \to 1$, gives $\delta\nu^{(2)}/\nu_0 \sim 10^{-76}$---still 57 orders of magnitude below $\sigma_y$. For $h_0 \sim \mathcal{O}(1)$, which occurs only in the strong-field region within a few Schwarzschild radii of a compact binary merger, linearized gravity breaks down and the tidal Hamiltonian [Eq.~(\ref{eq:H_tidal})] is no longer valid; moreover, tidal forces in that regime would disrupt the atom long before any measurement could be performed. No regime exists---weak field or strong---in which Mechanism~A produces a detectable signal in atoms.

\subsubsection{Polar molecules: evading the selection rule}
\label{sec:molecules}

The exact vanishing of the first-order energy shift (Sec.~\ref{sec:vanishing}) relies on $J = 0$ clock states. Heteronuclear diatomic molecules such as KRb, CaF, or SrF possess low-lying rotational states with $J \geq 1$, for which the Wigner--Eckart selection rule is satisfied: the $3j$ symbol $\bigl(\begin{smallmatrix}1 & 2 & 1\\0 & 0 & 0\end{smallmatrix}\bigr)$ is non-vanishing (since $1 + 2 + 1 = 4$ is even and the triangle inequality $|1 - 1| \leq 2 \leq 1 + 1$ holds). The tidal Hamiltonian therefore has non-zero diagonal matrix elements in $J = 1$ states, producing a first-order energy shift linear in $h_0$---a qualitative improvement over the quadratic scaling of atomic clock states.

The coupling also benefits from the nuclear mass scale. For internal molecular degrees of freedom, the tidal Hamiltonian [Eq.~(\ref{eq:H_tidal})] acts on the internuclear coordinate with the reduced mass $\mu = m_1 m_2/(m_1 + m_2)$ replacing the electron mass. The relevant length scale is the equilibrium bond length $r_0$, typically a few \AA. For KRb, $\mu \approx 26.7$~amu $= 4.4\times 10^{-26}$~kg and $r_0 = 4.0\times 10^{-10}$~m, giving
\begin{equation}
\mu\omega^2 r_0^2 = 4.4\times 10^{-26}\times(6.28\times 10^{-3})^2\times(4.0\times 10^{-10})^2 = 2.8\times 10^{-49}~\mathrm{J}
\end{equation}
at $f_{\mathrm{gw}} = 1$~mHz. For a $J = 1$, $M = 0$ rotational state in an optimally oriented tidal field, the diagonal matrix element of the angular operator yields an effective coefficient $C_{J=1} = 1/10$, computed from $\langle 1,0|n_y^2 - n_z^2|1,0\rangle = -2/5$ (where $\mathbf{n}$ is the molecular axis unit vector). The first-order shift at $h_0 = 10^{-21}$ is
\begin{equation}
\delta E^{(1)}_{\mathrm{mol}} = C_{J=1}\,\mu\omega^2 r_0^2\,h_0 = 2\times 10^{-52}~\mathrm{eV}.
\end{equation}
The rotational transition energy is $\Delta E_{\mathrm{rot}} = 2B \approx 2.2$~GHz~$\approx 9\times 10^{-6}$~eV (using $B = 1.114$~GHz for KRb), giving a fractional shift
\begin{equation}
\frac{\delta\nu_{\mathrm{mol}}}{\nu_0} \sim \frac{2\times 10^{-52}}{9\times 10^{-6}} \sim 2\times 10^{-47}.
\label{eq:mol_shift}
\end{equation}

This is $\sim$70 orders of magnitude larger than the atomic result ($\delta\nu^{(2)}/\nu_0 \sim 10^{-117}$). The gain comes from three sources: the shift is first-order rather than second-order in $h_0$ (gaining $\sim 10^{21}$ at $h_0 = 10^{-21}$), the reduced nuclear mass $\mu/m_e \sim 5\times 10^4$ replaces the electron mass, and there is no intermediate-state energy denominator suppression. However, the best demonstrated stability for molecular rotational clocks is $\sigma_y \sim 10^{-14}$~\cite{Kondov2019}, roughly five orders of magnitude worse than optical atomic clocks. Even with the most optimistic projected molecular clock stability of $\sigma_y \sim 10^{-16}$, the signal-to-noise deficit remains
\begin{equation}
\frac{\delta\nu_{\mathrm{mol}}/\nu_0}{\sigma_y} \sim \frac{2\times 10^{-47}}{10^{-16}} \sim 2\times 10^{-31}.
\end{equation}
Evading the $J = 0$ selection rule recovers 70 orders of magnitude relative to atomic clocks but leaves a deficit of $\sim$31 orders of magnitude, set by the tidal coupling strength at molecular scales. Polar molecules fare worse in the other two mechanisms: Mechanism~B sensitivity depends on clock stability $\sigma_y$, which is $\sim\!10^5$ times poorer for rotational clocks than for optical atomic clocks; Mechanism~C sensitivity scales with $k_{\mathrm{eff}} = n\omega_a/c$, which is $\sim\!10^5$ times smaller for GHz rotational transitions than for optical transitions. Molecules maximize the Mechanism~A coupling among all quantum systems considered, yet the remaining deficit is set by the coupling, not by clock performance.

\subsection{Mechanism B: External (center-of-mass) coupling}
\label{sec:mechanism_B}

Two quantum systems separated by baseline $L$ experience differential center-of-mass motion due to geodesic deviation. This manifests as two distinct effects.

\emph{Doppler shift.} For a monochromatic gravitational wave $h(t) = h_0\cos(\omega t)$, the relative velocity between two freely falling masses separated by baseline $L$ is $\delta v = \frac{1}{2}\dot{h}\,L$, with peak value $\delta v_{\mathrm{peak}} = \frac{1}{2}\omega h_0 L$. The resulting Doppler shift on an optical carrier of frequency $\nu_0$ is $\delta\nu_{\mathrm{Doppler}} = \nu_0\,\delta v/c$, giving a peak fractional frequency shift
\begin{equation}
\frac{\delta\nu_{\mathrm{Doppler}}}{\nu_0} = \frac{\omega h_0 L}{2c}.
\label{eq:doppler}
\end{equation}
The minimum detectable strain is determined by the clock's fractional frequency stability. We characterize this by the Allan deviation $\sigma_y(\tau)$, the standard measure of fractional frequency fluctuations $y(t) \equiv \Delta\nu(t)/\nu_0$ averaged over integration time $\tau$. A signal is detectable when the induced shift exceeds $\sigma_y$:
\begin{equation}
h_{\mathrm{min}}^{\mathrm{Doppler}} = \frac{2c\,\sigma_y}{\omega L}.
\label{eq:hmin_doppler}
\end{equation}
Two distinct clock performance metrics are relevant. The \emph{frequency stability} $\sigma_y(\tau) = 3.5\times 10^{-16}/\sqrt{\tau/\mathrm{s}}$~\cite{Marshall2025} determines the noise floor for measurements at integration time $\tau$ and sets the instantaneous sensitivity to time-varying signals. The \emph{systematic uncertainty} $\sigma_y^{\mathrm{sys}} = 5.5\times 10^{-19}$~\cite{Marshall2025} represents the ultimate accuracy floor, reached after integration times of $\tau_{\mathrm{floor}} \sim (\sigma_y(1\,\mathrm{s})/\sigma_y^{\mathrm{sys}})^2 \sim 4\times 10^{5}$~s ($\sim$5~days). We use the appropriate metric for each calculation: the stability for Doppler sensitivity (which requires detecting a time-varying signal at the gravitational wave period) and the systematic uncertainty for differential proper time (which benefits from long averaging).

Using $\sigma_y(\tau = 1\,\mathrm{s}) = 3.5\times 10^{-16}$, with $L = 2.5$~Gm (LISA arm length) and $f = 1$~mHz:
\begin{equation}
h_{\mathrm{min}}^{\mathrm{Doppler}} = \frac{2\times(3\times 10^8)\times(3.5\times 10^{-16})}{(6.28\times 10^{-3})\times(2.5\times 10^9)} = 1.3\times 10^{-14},
\label{eq:hmin_doppler_num}
\end{equation}
a factor of $1.3\times 10^{7}$ above the LISA sensitivity target of $h \sim 10^{-21}$.

\emph{Differential proper time.} Two clocks at different positions accumulate different proper times:
\begin{equation}
\frac{\delta\tau}{\tau} = \frac{1}{2}h\sin\!\left(\frac{\omega L}{c}\right).
\label{eq:time_dilation}
\end{equation}
For LISA, $\omega L/c = 0.052$ at 1~mHz (marginally satisfying $\omega L/c \ll 1$). To provide a generous upper bound, we assume $N = 10$ clocks at distinct locations along the LISA constellation arms, yielding $N_{\mathrm{eff}} = N - 1 = 9$ independent baselines. With the systematic uncertainty $\sigma_y^{\mathrm{sys}} = 5.5\times 10^{-19}$~\cite{Marshall2025}:
\begin{align}
h_{\mathrm{min}}^{\mathrm{dilation}} &= \frac{\sigma_y^{\mathrm{sys}}}{\frac{1}{2}\sin(\omega L/c)\sqrt{N_{\mathrm{eff}}}} \nonumber\\
&= \frac{5.5\times 10^{-19}}{0.5\times 0.052\times\sqrt{9}} = 7.1\times 10^{-18},
\label{eq:hmin_dilation}
\end{align}
a factor of $\sim 10^4$ above the LISA target.

These results use the most optimistic published clock parameters~\cite{Marshall2025}. Even with arbitrarily perfect clocks ($\sigma_y \to 0$), the Doppler and time-dilation approaches are limited by different fundamental constraints (atom number, coherence time, integration time), none of which can bridge the remaining sensitivity gap.

\subsection{Mechanism C: Propagation coupling}
\label{sec:mechanism_C}

Light traveling between two freely falling test masses accumulates a phase shift proportional to the integrated metric perturbation along the photon path:
\begin{equation}
\delta\varphi = \frac{2\pi\nu}{c}\int_0^L h(t,x)\,dx \approx \frac{\pi\nu h L}{c} \quad (L \ll \lambda_{\mathrm{gw}}).
\label{eq:propagation}
\end{equation}
The transducer gain---the phase shift per unit strain---is
\begin{equation}
G_{\mathrm{interf}} = \frac{\delta\varphi}{h} = \frac{\pi\nu L}{c} = 7.38\times 10^{15}~\mathrm{rad},
\label{eq:G_interf}
\end{equation}
for LISA parameters ($\nu = 2.82\times 10^{14}$~Hz, $L = 2.5$~Gm). This enormous transducer gain arises from the $\sim 2.35\times 10^{15}$ optical wavelengths spanning the arm length and is the fundamental reason that laser interferometry is the method of choice for gravitational wave detection.

\subsection{The transducer gain hierarchy}
\label{sec:hierarchy}

Comparing the three coupling mechanisms in terms of the dimensionless observable produced per unit strain:
\begin{align}
G_{\mathrm{interf}} &= \frac{\pi\nu L}{c} = 7.38\times 10^{15}, \label{eq:G1}\\
G_{\mathrm{Doppler}} &= \frac{\omega L}{2c} = 2.62\times 10^{-2}, \label{eq:G2}\\
G_A &= 2.4\times 10^{-20}. \label{eq:G3}
\end{align}
The ratio $G_{\mathrm{interf}}/G_A = 3.1\times 10^{35}$, combined with the exact vanishing of the first-order clock shift [Eq.~(\ref{eq:clock_vanish})] and the $h_0^2$ suppression at second order [Eq.~(\ref{eq:second_order_numerical})], constitutes the \emph{no-go result} for direct quantum detection of gravitational waves via internal atomic coupling. These three barriers are distinct; circumventing one does not alleviate the others.

The physical origin of this hierarchy is clear. Interferometry reads out the gravitational-wave phase imprinted on light propagating over a macroscopic baseline ($L \sim 10^9$~m), whereas internal atomic coupling reads out the tidal distortion of a bound system of atomic size ($a_0 \sim 10^{-11}$~m). The ratio combines the geometric lever arm $(L/a_0)^2 \sim 10^{40}$, the frequency ratio $\nu_{\mathrm{optical}}/\omega_{\mathrm{gw}} \sim 10^{17}$, and the inverse atomic velocity parameter, partially canceling but still yielding a ratio of $\sim 10^{35}$.

Even the Doppler approach (Mechanism B), which exploits the macroscopic baseline $L$, falls short by $\sim 10^{7}$ because it is limited by the fractional frequency stability $\sigma_y$ rather than by the optical phase sensitivity. These deficits are set by the coupling mechanism and exceed any projected quantum enhancement.

\subsection{Gauge invariance and the meaning of the comparison}
\label{sec:gauge}

The three transducer gains in Eqs.~(\ref{eq:G1})--(\ref{eq:G3}) are built from gauge-invariant quantities, and the classification does not depend on a choice of coordinates. For Mechanisms~A and~B the coupling is governed by the electric part of the Riemann tensor $R^i{}_{0j0}$ evaluated on the observer's worldline [Eq.~(\ref{eq:H_tidal})], a curvature scalar rather than a coordinate artifact, obtained above by passing from the TT-gauge metric to the proper detector frame [Eqs.~(\ref{eq:Rx})--(\ref{eq:Rz})]. For Mechanism~C the invariant is instead the phase accumulated by light along its null path between the worldlines [Eq.~(\ref{eq:propagation})]; although written in TT gauge, the measured round-trip phase does not depend on that choice. In all three cases the two descriptions give the same physics. What distinguishes the mechanisms is not where the effect is said to reside but which gauge-invariant observable is read out: an internal energy-level shift (Mechanism~A), a fractional-frequency or proper-time ratio between separated worldlines (Mechanism~B), or the optical phase accumulated between worldlines (Mechanism~C).

For an interferometer this requires care. Whether one says the test masses move (proper detector frame) or the light propagation is modified (TT gauge) is a gauge-dependent statement; the round-trip phase that is actually measured is the same gauge-invariant quantity in either description, and $G_{\mathrm{interf}}$ [Eq.~(\ref{eq:G_interf})] is that quantity. The classification therefore rests on the readout channel, not on a gauge-dependent partition of the signal. Quantities that might appear to constitute separate mechanisms---clock comparison, laser phase noise, relativistic reference-frame corrections---are not additional channels: a clock comparison is the readout of Mechanism~B, laser phase noise is a common-mode contribution---suppressed by time-delay interferometry in unequal-arm configurations such as LISA (Sec.~\ref{sec:tdi_suppression})---rather than a coupling to the wave, and reference-frame corrections are the coordinate expression of the same worldline curvature. Unified treatments of clocks and atom interferometers reach the same conclusion, the two differing by whether the atoms are confined rather than by the underlying coupling~\cite{Norcia2017,Schaffrath2025}.

It follows that the three gains are not directly commensurable as written: $G_A$ is a dimensionless ratio of an energy shift to a level spacing, $G_{\mathrm{Doppler}}$ a fractional-frequency response, and $G_{\mathrm{interf}}$ a phase per unit strain. They are transducer gains into different readout channels, and become comparable only once each is paired with the noise floor of its channel. The physically homogeneous figure of merit is the minimum detectable strain, dimensionless and of the same kind for all three; we collect this comparison in Sec.~\ref{sec:noise} (Table~\ref{tab:hmin}). The hierarchy of Eqs.~(\ref{eq:G1})--(\ref{eq:G3}) should be read accordingly: it records how strongly each channel transduces strain, not a direct comparison of sensitivities.

\section{Direct Quantum Detection in Current Schemes}
\label{sec:direct}

We now apply the framework of Sec.~\ref{sec:coupling} to concrete direct-detection schemes that have appeared in the literature, using the parameters published for each, to see how the coupling mechanism plays out in real cases.

\subsection{Kolkowitz et al.: Optical lattice clock comparison}
\label{sec:kolkowitz}

Kolkowitz, Pikovski, Langellier, Lukin, Walsworth, and Ye~\cite{Kolkowitz2016} proposed detecting gravitational waves through the differential frequency shift between two optical lattice clocks separated by a baseline $L$, an approach that builds on the use of optical atomic transitions as ultra-stable frequency and phase references~\cite{Hollberg2017}. The mechanism relies on the Doppler shift (Mechanism B): the gravitational wave induces differential motion between the lattice sites, producing a frequency shift in the exchanged laser light.

The strain sensitivity is determined by the clock comparison uncertainty:
\begin{equation}
h_{\mathrm{min}} = \frac{2c\,\sigma_y}{\omega_{\mathrm{gw}} L},
\end{equation}
where $\sigma_y$ is the fractional frequency instability achievable with the clock comparison. The proposal envisions $N = 10^6$ atoms per lattice with interrogation time $\tau$. The quantum projection noise gives $\sigma_y(\tau) = 1/(\omega_0\sqrt{N}\,\tau)$, where $\omega_0$ is the clock transition frequency. Over many cycles, the instability averages as $\sigma_y(T_{\mathrm{obs}}) = \sigma_y(\tau)/\sqrt{T_{\mathrm{obs}}/\tau}$.

Using the parameters stated in~\cite{Kolkowitz2016}: $\omega_0 = 2\pi\times 4.29\times 10^{14}$~Hz (${}^{87}$Sr clock transition), $N = 10^6$, $\tau = 1$~s, $L = 2.5$~Gm, and $f_{\mathrm{gw}} = 1$~mHz, the achievable instability per cycle is $\sigma_y(\tau) = 3.7\times 10^{-19}$. However, this assumes the lattice sites can be maintained at the required separation with picometer-level stability over interplanetary distances, which is precisely the engineering challenge that LISA solves through heterodyne laser interferometry with TDI.

The fundamental limitation is that clock-based detection relies on Mechanism B (center-of-mass Doppler shift), with transducer gain $G_{\mathrm{Doppler}} = \omega L/(2c) = 2.6\times 10^{-2}$, while LISA exploits Mechanism C (light propagation), with $G_{\mathrm{interf}} = 7.4\times 10^{15}$. The ratio of $\sim 10^{17}$ means the clock must achieve fractional frequency stability $10^{17}$ times better than the LISA interferometer's phase sensitivity to be competitive---a gap set by the coupling mechanism, not by the quantum sensor.

\subsection{Loeb and Maoz: Atomic clock time dilation}
\label{sec:loeb}

Loeb and Maoz~\cite{Loeb2016} proposed using atomic clocks to detect the differential time dilation produced by gravitational waves [Mechanism B, Eq.~(\ref{eq:time_dilation})]. As computed in Sec.~\ref{sec:mechanism_B}, even with the best current clock performance ($\sigma_y = 5.5\times 10^{-19}$~\cite{Marshall2025}), the minimum detectable strain is $h_{\mathrm{min}} = 7.1\times 10^{-18}$---a factor of $\sim 10^{4}$ above the LISA target.

To reach LISA sensitivity through time dilation alone would require $\sigma_y \sim 5\times 10^{-23}$, which is four orders of magnitude beyond the current state of the art and three orders of magnitude beyond any projected clock capability. This gap is fundamental: it reflects the transducer gain hierarchy (Sec.~\ref{sec:hierarchy}), in which the time-dilation signal scales as $\omega L/c$ while the interferometric phase signal scales as $\nu_{\mathrm{optical}} L/c$.

\section{Noise Architecture and Hybrid Quantum Enhancement}
\label{sec:noise}

Given the transducer gain hierarchy established in Sec.~\ref{sec:coupling}, the practical question becomes whether quantum sensors can \emph{enhance} the sensitivity of interferometric detectors by reducing specific noise contributions. The answer depends critically on the noise architecture of the detector.

Before turning to that question, Table~\ref{tab:hmin} collects the three mechanisms on the only common footing that is physically meaningful---the minimum detectable strain, dimensionless and of the same kind across rows---rather than the transducer gains of Eqs.~(\ref{eq:G1})--(\ref{eq:G3}), which are not directly commensurable (Sec.~\ref{sec:gauge}). Mechanisms~A and~B fall short of the strain levels of interest by the coupling mechanism itself; only Mechanism~C reaches them, and does so as a strain spectral density rather than a single threshold.

\begin{table}[htbp]
\centering
\caption{\label{tab:hmin}Minimum detectable strain for the three coupling mechanisms, evaluated for LISA parameters at $f = 1$~mHz. The entries are placed on a common strain footing (Sec.~\ref{sec:gauge}); the transducer gains themselves [Eqs.~(\ref{eq:G1})--(\ref{eq:G3})] are not directly comparable. Mechanism~A is quadratic in strain (first order vanishes for $J=0$), so its entry is the fractional clock shift evaluated at a fixed reference strain. For reference, the LISA design target is $h \sim 10^{-21}$.}
{\small
\begin{tabular}{@{}llll@{}}
\hline
Mechanism & Observable read out & Limited by & $h_{\mathrm{min}}$ \\
\hline
A: internal tidal & energy-level shift & coupling structure$^{a}$ & $\delta\nu/\nu_0 \sim 10^{-117}$\,$^{b}$ \\
B: Doppler & fractional frequency $\delta\nu/\nu_0$ & clock stability $\sigma_y$ & $1.3\times 10^{-14}$ \\
B: proper time & proper-time ratio $\delta\tau/\tau$ & clock systematics $\sigma_y^{\mathrm{sys}}$ & $7.1\times 10^{-18}$ \\
C: propagation & optical phase $\delta\varphi$ & shot / OMS noise & reaches design sensitivity$^{c}$ \\
\hline
\end{tabular}}
\begin{flushleft}
\footnotesize
$^{a}$ First-order shift vanishes for $J=0$ states; the leading shift is second order, $\propto h_0^2$.\\
$^{b}$ Fractional shift at $h_0 = 10^{-21}$; being quadratic in strain, it is not a linear threshold.\\
$^{c}$ Expressed natively as a strain spectral density $\sqrt{S_h(f)}$ [Eq.~(\ref{eq:Sh})], not as a single strain value; reaches the design target above.
\end{flushleft}
\end{table}

\subsection{Enhancement formula}
\label{sec:universal}

For any noise source contributing fraction $\beta$ of the total noise power $S_h$, a quantum sensor that reduces this contribution by factor $\eta$ (so the noise PSD becomes $S_i/\eta^2$) yields a strain sensitivity enhancement
\begin{equation}
\mathcal{E} = \frac{1}{\sqrt{1 - \beta\!\left(1 - 1/\eta^2\right)}}.
\label{eq:enhancement}
\end{equation}
This formula follows from the definition $S_h^{\mathrm{new}} = S_h[1 - \beta(1 - 1/\eta^2)]$ and $\mathcal{E} = \sqrt{S_h^{\mathrm{old}}/S_h^{\mathrm{new}}}$.

In the limit of arbitrarily perfect quantum sensors ($\eta \to \infty$):
\begin{equation}
\mathcal{E}_{\mathrm{max}} = \frac{1}{\sqrt{1-\beta}}.
\label{eq:E_max}
\end{equation}
This identity constrains all hybrid enhancement strategies: \emph{the ceiling on quantum enhancement is set by the noise architecture; the quantum technology determines how close to that ceiling the detector operates.} The quantity $\beta$ depends on both the detector's noise budget and the set of noise sources that available quantum technologies can address; it is a practical figure of merit, not a fundamental constant. For $\beta \ll 1$, this gives $\mathcal{E}_{\mathrm{max}} \approx 1 + \beta/2$: the enhancement is bounded by half the noise fraction, regardless of how much the quantum sensor improves.

For multiple independent quantum channels targeting distinct noise sources:
\begin{equation}
\mathcal{E}_{\mathrm{total}}(f) = \frac{1}{\sqrt{1 - \sum_i \beta_i(f)\left(1 - 1/\eta_i^2\right)}},
\label{eq:combined}
\end{equation}
where $\beta_i(f)$ is the frequency-dependent fraction of total noise power from each quantum-accessible source.

\subsection{LISA sensitivity and Time-Delay Interferometry}
\label{sec:LISA_TDI}

The LISA sensitivity is determined by two statistically independent noise sources combined through TDI transfer functions~\cite{Robson2019,LISARedBook2024}:

\begin{equation}
S_h(f) = \frac{10}{3L^2}\left[S_{\mathrm{OMS}} + \frac{2(1+\cos^2(f/f_*))}{(2\pi f)^4}\,S_a\right]\!\left[1 + 0.6\left(\frac{f}{f_*}\right)^{\!2}\right],
\label{eq:Sh}
\end{equation}

where $f_* = c/(2\pi L) = 19.1$~mHz is the transfer frequency, and the noise PSDs are~\cite{LISARedBook2024}
\begin{align}
S_a &= (3\!\times\! 10^{-15})^2\!\left[1 + \left(\frac{0.4\,\mathrm{mHz}}{f}\right)^{\!2}\right] \nonumber\\
&\quad\times\left[1 + \left(\frac{f}{8\,\mathrm{mHz}}\right)^{\!4}\right]\frac{\mathrm{m}^2}{\mathrm{s}^4\,\mathrm{Hz}}, \label{eq:Sa}\\
S_{\mathrm{OMS}} &= (15\!\times\! 10^{-12})^2\!\left[1 + \left(\frac{2\,\mathrm{mHz}}{f}\right)^{\!4}\right]\frac{\mathrm{m}^2}{\mathrm{Hz}}. \label{eq:SOMS}
\end{align}

There is no separate laser frequency noise term in Eq.~(\ref{eq:Sh}). This is by design: TDI suppresses laser phase noise below the OMS floor. This fact is central to the quantum enhancement analysis and is often overlooked in proposals for atomic clock improvements to LISA.

\begin{figure*}[htbp]
\centering
\includegraphics[width=\textwidth]{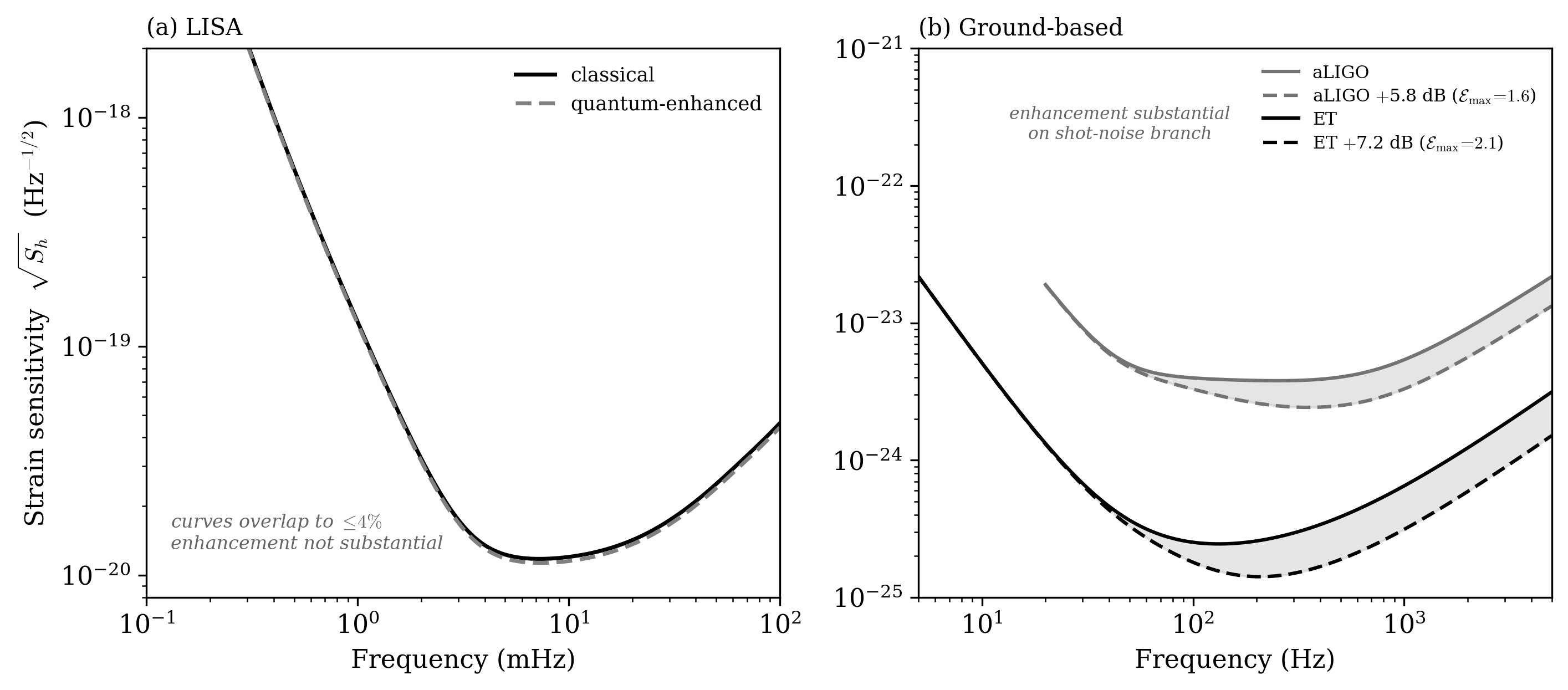}
\caption{\label{fig:enhancement}Strain sensitivity before and after quantum enhancement, contrasting the two noise-architecture regimes. \textbf{(a)} LISA: classical sensitivity from Eq.~(\ref{eq:Sh}) and the quantum-enhanced curve obtained by applying the combined enhancement $\mathcal{E}(f)$ of Table~\ref{tab:enhancement}. The two curves overlap to within the ceiling $\mathcal{E}\leq1.04$ set by $\beta\approx0.09$; the enhancement is not substantial in any band. \textbf{(b)} Ground-based detectors: design sensitivities of Advanced LIGO (aLIGOZeroDetHighPower analytic fit~\cite{SamajdarArun2017}) and the Einstein Telescope (ET-B fit~\cite{Mishra2010}), with quantum-enhanced curves at the effective squeezing levels of Table~\ref{tab:ground} (5.8~dB and 7.2~dB). The two panels use different vertical axes---strain sensitivity in both, but spanning the detectors' respective scales---because a few-percent and a factor-of-two enhancement cannot be displayed legibly on a common range; this difference is itself the point. The enhancement shown in panel~(b) is that of the shot-noise-dominated regime evaluated in Table~\ref{tab:ground} ($f\gtrsim200$~Hz); the frequency-dependent shape of the transition is illustrative, while the peak values $\mathcal{E}_{\max}=1.6$ and $2.1$ are those of Table~\ref{tab:ground}. The broadband enhancement realized in LIGO O4 with frequency-dependent squeezing extends to lower frequencies through simultaneous reduction of radiation-pressure noise~\cite{Jia2024,ZhangMiao2025}, a regime not captured by the single-quadrature, high-frequency evaluation shown here.}
\end{figure*}

The structure of this budget determines directly why quantum enhancement is ineffective for LISA, and because the reasoning separates three physically distinct noise contributions, we state it explicitly here before quantifying each in turn. First, laser frequency noise---the largest raw contribution---does not limit the enhanced sensitivity: second-generation TDI suppresses it to a fraction $\beta_{\mathrm{laser}} = 5\times 10^{-5}$ of the residual budget (Sec.~\ref{sec:tdi_suppression}), so it is removed by design rather than remaining as a residual obstacle. Second, below $\sim$4~mHz the residual budget is dominated by test-mass acceleration noise [$S_a$, Eq.~(\ref{eq:Sa})], which is classical in origin---principally Brownian force noise from residual gas and actuation-related disturbances~\cite{LISARedBook2024}---and is not addressable by quantum readout. Third, above $\sim$4~mHz the budget is dominated by the optical metrology system [$S_{\mathrm{OMS}}$, Eq.~(\ref{eq:SOMS})], of which photon shot noise, the only genuinely quantum-limited component, is a minor part---bounded to $\lesssim 15\%$ by the received optical power (Sec.~\ref{sec:oms_budget}). The consequence is that in \emph{neither} frequency regime does quantum-accessible noise represent a substantial fraction of the total: below 4~mHz because classical acceleration noise dominates, and above 4~mHz because shot noise is a minor part of the OMS. Taken across the band, the quantum-accessible fraction is $\beta \approx 0.09$, giving a ceiling $\mathcal{E}_{\mathrm{max}} = 1/\sqrt{1-\beta} \approx 1.04$ irrespective of how ideal the quantum sensor is. This is the physical origin of the near-coincidence of the classical and quantum-enhanced curves in Fig.~\ref{fig:enhancement}(a); the following subsections quantify each of the three contributions.

\subsubsection{TDI suppression of laser frequency noise}
\label{sec:tdi_suppression}

The LISA laser, pre-stabilized to a Fabry--P\'{e}rot cavity, has a frequency noise of
$\sqrt{S_\nu} \approx 30$~Hz/$\sqrt{\mathrm{Hz}}$ in the millihertz band, corresponding to fractional instability~\cite{LISARedBook2024}
\begin{equation}
\sigma_y^{\mathrm{laser}} = \frac{\sqrt{S_\nu}}{\nu} = \frac{30}{2.82\times 10^{14}} = 1.06\times 10^{-13}.
\label{eq:sigma_laser}
\end{equation}
Without suppression, this would dominate the noise budget by many orders of magnitude.

Second-generation TDI, combining the six one-way phase measurements with appropriate time delays, synthesizes virtual equal-arm interferometers~\cite{Tinto2014,Tinto2021}. With arm length knowledge $\delta L \sim 1$~m from inter-spacecraft ranging, the laser noise residual in the TDI output is
\begin{equation}
\sqrt{S_x^{\mathrm{TDI}}} = \sigma_y^{\mathrm{laser}} \times \delta L = 1.06\times 10^{-13} \times 1~\mathrm{m} = 0.106~\frac{\mathrm{pm}}{\sqrt{\mathrm{Hz}}}.
\label{eq:tdi_residual}
\end{equation}
The resulting noise power fraction in the OMS budget is
\begin{equation}
\beta_{\mathrm{laser}} = \left(\frac{0.106}{15}\right)^{\!2} = 5.0\times 10^{-5}.
\label{eq:beta_laser}
\end{equation}
This is 0.005\% of the OMS noise power. The estimate scales as $\delta L^2$ with the inter-spacecraft ranging knowledge; over the plausible range $\delta L \sim 0.3$--$3$~m it stays within $\beta_{\mathrm{laser}} \sim 5\times 10^{-6}$--$5\times 10^{-4}$, so TDI reduces laser frequency noise to negligibility in all cases.

An atomic clock improving the laser reference by a factor $\eta = 250$ (from $10^{-13}$ to $4\times 10^{-16}$) would reduce this already negligible contribution by an additional $\eta^2 = 62{,}500$, yielding
\begin{equation}
\mathcal{E}_{\mathrm{clock}}^{\mathrm{laser}} = \frac{1}{\sqrt{1 - 5\times 10^{-5}}} \approx 1.000025.
\end{equation}
\emph{Atomic clock stabilization of the LISA laser provides no measurable improvement in strain sensitivity.} The TDI algorithm has already solved the laser noise problem by algorithmic rather than hardware means.

\subsubsection{OMS sub-budget decomposition}
\label{sec:oms_budget}

The 15~pm/$\sqrt{\mathrm{Hz}}$ OMS requirement is an RSS allocation encompassing all measurement noise sources in the heterodyne interferometer on each optical bench~\cite{LISARedBook2024}. What matters for quantum enhancement is not the detailed partitioning among these sources, but the fraction that is quantum-limited, summarized in Table~\ref{tab:oms_budget}.

\begin{table}
\caption{\label{tab:oms_budget}Quantum-accessible versus classical fraction of the OMS noise budget. The detailed per-component partitioning of the LISA OMS budget is internal to the LISA Consortium; what matters for quantum enhancement is that photon shot noise---the only genuinely quantum-limited contribution---is bounded to $\lesssim 15\%$ of the OMS budget by the received optical power ($\sim$700~pW; see text). This is corroborated by the in-flight performance of the LISA Pathfinder optical metrology system, in which phasemeter readout noise dominated and shot noise was not a limiting contributor~\cite{Armano2021,Armano2022}. USO clock jitter contributes at the $\sim$1\% level and is addressable by an atomic clock; since shot noise is bounded to $\lesssim 15\%$, the classical contributions make up the remaining $\gtrsim 85\%$ and lie outside the reach of quantum techniques.}
\begin{tabular}{@{}lcc@{}}
Contribution & Nature & Fraction of OMS budget \\
\hline
Photon shot noise & Quantum & $\lesssim 15\%$ \\
USO clock jitter & Quantum-addressable & $\sim 1\%$ \\
Phasemeter, pathlength, & Classical & remainder \\
\quad tilt-to-length, thermal, stray light & & \\
\end{tabular}

\end{table}

The structure of this budget is fundamentally different from ground-based detectors. Classical noise sources---phasemeter electronics, optical bench thermal and mechanical fluctuations, and tilt-to-length coupling---collectively dominate the OMS budget. The only contribution accessible to squeezing is photon shot noise, and USO clock jitter is addressable by an atomic clock. This asymmetry is the physical reason that quantum enhancement of LISA is marginal.

To assess robustness against uncertainty in the OMS decomposition, we note that the shot noise fraction is bounded by the optical design: with $\sim$700~pW received power and the heterodyne readout architecture, the shot noise contribution cannot exceed $\sim$15\% of the OMS budget under any reasonable allocation. This is consistent with in-flight data from LISA Pathfinder, where the OMS noise was dominated by phasemeter readout and laser frequency noise, with shot noise not a limiting contributor under nominal operating conditions~\cite{Armano2021,Armano2022}. Even at the upper bound $\beta_{\mathrm{shot}} = 0.15$, the maximum squeezed vacuum enhancement would be $\mathcal{E}_{\mathrm{sq}}^{\mathrm{max}} = 1/\sqrt{1 - 0.15} = 1.08$, which remains marginal. The conclusion that LISA's classical noise architecture limits quantum enhancement to $\mathcal{E} \lesssim 1.05$ is therefore robust against factor-of-two uncertainties in the OMS sub-budget.

\subsection{LISA quantum enhancement channels}
\label{sec:LISA_channels}

\emph{Channel 1: Squeezed vacuum injection.} Squeezed vacuum replaces the vacuum fluctuations entering the readout port, reducing the shot noise PSD by factor $\xi_{\mathrm{eff}} = 10^{r_{\mathrm{eff}}/10}$, where $r_{\mathrm{eff}}$ is the effective squeezing in decibels. With 8.3~dB effective squeezing (accounting for optical losses $\mathcal{L} \approx 0.12$ via $V_{\mathrm{eff}} = (1-\mathcal{L})/\xi_{\mathrm{raw}} + \mathcal{L}$), $\xi_{\mathrm{eff}} = 6.76$. The shot noise fraction $\beta_{\mathrm{shot}} \approx 0.08$ gives
\begin{equation}
\mathcal{E}_{\mathrm{sq}} = \frac{1}{\sqrt{1 - 0.08\times(1 - 1/6.76)}} = 1.035.
\label{eq:E_squeeze}
\end{equation}
Even with infinite squeezing, $\mathcal{E}_{\mathrm{sq}}^{\mathrm{max}} = 1/\sqrt{1 - 0.08} = 1.043$.

\emph{Channel 2: Atomic clock USO replacement.} The LISA phasemeter timestamps the heterodyne beatnote using a local ultra-stable oscillator with $\sigma_y \sim 10^{-13}$. An atomic clock with $\sigma_y \sim 4\times 10^{-16}$ would improve this by $\eta_{\mathrm{clock}} \approx 250$. With $\beta_{\mathrm{USO}} \approx 0.012$:
\begin{equation}
\mathcal{E}_{\mathrm{clock}} = \frac{1}{\sqrt{1 - 0.012\times(1 - 1/250^2)}} = 1.006.
\label{eq:E_clock}
\end{equation}

\emph{Channel 3: Quantum magnetometry.} The LISA test masses experience residual magnetic forces characterized by the fraction $\gamma_{\mathrm{mag}} = 0.01$--$0.055$ of the acceleration noise budget~\cite{LISARedBook2024}. The magnetic-field-independent SERF magnetometer proposed by Dikopoltsev, Levy, and Katz~\cite{Dikopoltsev2025}, which projects a fundamental sensitivity of $\sim 10$~aT$\sqrt{\mathrm{cm}^3/\mathrm{Hz}}$ at geomagnetic fields (a theoretical limit for feasible experimental conditions, not yet demonstrated), extends the operational range of SERF magnetometry beyond the low-field regime where subfemtotesla sensitivities have been achieved~\cite{Kominis2003}. In the acceleration-dominated regime:
\begin{equation}
\mathcal{E}_{\mathrm{mag}} = \frac{1}{\sqrt{1 - 0.03\times(1 - 1/100)}} = 1.015.
\label{eq:E_mag}
\end{equation}

The fraction $\gamma_{\mathrm{mag}}$ is itself a range; propagating $\gamma_{\mathrm{mag}} = 0.01$--$0.055$ through Eq.~(\ref{eq:E_mag}) gives $\mathcal{E}_{\mathrm{mag}} \approx 1.005$--$1.028$ in the acceleration-dominated band, with the value above corresponding to the midpoint. The reduction factor $\eta = 100$ assumed here is a technological parameter, not a derived one. The fraction $\gamma_{\mathrm{mag}}$ is fixed by the noise architecture, but how much of it is actually subtractable is governed by the magnetic-force-to-acceleration transfer function, the spatial coherence between the magnetometer location and the test mass, and the calibration noise of the subtraction model---quantities set by the specific implementation rather than by linearized general relativity or non-relativistic quantum mechanics. A realistic subtraction model therefore depends on the detailed magnetic coupling of a given mission and falls in the technology-dependent category; we do not develop it here.

\emph{Combined enhancement.} Because the three channels target statistically independent noise sources, their effects combine via Eq.~(\ref{eq:combined}). The frequency-dependent results are shown in Table~\ref{tab:enhancement}. The maximum combined enhancement is $\mathcal{E}_{\mathrm{total}} \approx 1.04$ in the OMS-dominated regime ($f \gtrsim 5$~mHz), corresponding to a survey volume increase of $\mathcal{E}^3 \approx 1.13$ (13\%).

\begin{table}
\caption{\label{tab:enhancement}Frequency-dependent strain sensitivity enhancement $\mathcal{E}(f)$ from quantum technologies applied to LISA. $f_a$ and $f_{\mathrm{OMS}}$ are the fractional contributions of acceleration noise and OMS noise to the total.}
\begin{tabular}{@{}ccccccc@{}}
$f$ (mHz) & $f_a$ & $f_{\mathrm{OMS}}$ & $\mathcal{E}_{\mathrm{sq}}$ & $\mathcal{E}_{\mathrm{clk}}$ & $\mathcal{E}_{\mathrm{mag}}$ & $\mathcal{E}_{\mathrm{tot}}$ \\
\hline
0.1 & 0.991 & 0.009 & 1.000 & 1.000 & 1.015 & 1.015 \\
0.3 & 0.947 & 0.053 & 1.002 & 1.000 & 1.014 & 1.017 \\
1.0 & 0.875 & 0.125 & 1.004 & 1.001 & 1.013 & 1.019 \\
3.0 & 0.520 & 0.480 & 1.017 & 1.003 & 1.008 & 1.029 \\
5.0 & 0.152 & 0.848 & 1.031 & 1.005 & 1.002 & 1.039 \\
10 & 0.030 & 0.970 & 1.036 & 1.006 & 1.000 & 1.043 \\
30 & 0.012 & 0.988 & 1.037 & 1.006 & 1.000 & 1.043 \\
100 & 0.015 & 0.985 & 1.036 & 1.006 & 1.000 & 1.043 \\
\end{tabular}

\end{table}

For comparison, a 10\% reduction in the dominant classical noise sources (optical bench pathlength, tilt-to-length coupling, phasemeter electronics) would provide comparable or greater strain improvement than the entire quantum enhancement budget. The engineering effort required for space-qualified quantum sensors must be weighed against this alternative.

These figures are specific to the approved LISA baseline~\cite{LISARedBook2024}. The value $\beta \approx 0.09$, and the ceiling $\mathcal{E} \approx 1.04$ that follows from it, are set by LISA's established hardware budget, not by any limitation of clocks or quantum sensors in space generally; the enhancement formula [Eq.~(\ref{eq:enhancement})] applies to any architecture through its own $\beta$. A mission adopting a different noise budget---employing clocks for calibration, redundancy, or clock-noise transfer rather than direct laser stabilization, or a detector topology departing from the conventional layout~\cite{Krenn2025}---would have a different quantum-accessible fraction and a correspondingly different ceiling.

\section{Atom Interferometric Detectors}
\label{sec:atom_interf}

Atom interferometric detectors represent a qualitatively different application of quantum sensors to gravitational wave detection and merit separate treatment. Unlike the hybrid enhancement channels of Sec.~\ref{sec:LISA_channels}, which improve existing classical detectors at the margins, atom interferometers propose an entirely new detector architecture in which quantum mechanics is integral to both signal transduction and noise properties.

\subsection{Operating principle}
\label{sec:ai_principle}

A light-pulse atom interferometer uses a sequence of laser pulses ($\pi/2$--$\pi$--$\pi/2$, separated by interrogation time $T$) to split, redirect, and recombine the de~Broglie wave of an atom in free fall. The atom accumulates phase along two spatially separated paths; the differential phase encodes the acceleration:
\begin{equation}
\Phi = k_{\mathrm{eff}} \cdot a \cdot T^2,
\label{eq:AI_phase}
\end{equation}
where $k_{\mathrm{eff}} = n \times 2\pi/\lambda$ is the effective wavevector enhanced by a factor $n$ through large momentum transfer (LMT) techniques.

A single atom interferometer cannot distinguish a gravitational wave from other inertial effects. The use of atom interferometry for gravitational wave detection was first proposed in the atomic gravitational wave interferometric sensor (AGIS) concept~\cite{Dimopoulos2008}. The essential idea of Graham et al.~\cite{Graham2013} is to operate two atom interferometers separated by baseline $L$, interrogated by the same laser beam using single-photon clock transitions (${}^{87}$Sr at 698~nm or ${}^{171}$Yb at 578~nm). The differential phase shift [cf.\ Eq.~(1) of Ref.~\cite{Graham2013}] is sensitive to the gravitational wave strain through the light travel time across the baseline:
\begin{equation}
\Delta\Phi_{\mathrm{gw}} = n\,\frac{\omega_a}{c}\, h\, L\, \sin^2\!\!\left(\frac{\omega_{\mathrm{gw}} T}{2}\right)\frac{4}{(\omega_{\mathrm{gw}}T)^2},
\label{eq:diff_phase}
\end{equation}
where $n$ is the large momentum transfer (LMT) order and $\omega_a$ is the atomic transition angular frequency. The factor $n\omega_a/c = n k_{\mathrm{laser}}$ is the effective wave vector, and the $\sin^2(\omega_{\mathrm{gw}} T/2) \cdot 4/(\omega_{\mathrm{gw}}T)^2$ is the atom interferometer transfer function $\mathcal{T}(f)$. This function is maximum at $f = 0$ (where $\mathcal{T} \to 1$) and decreases monotonically, falling to $\mathcal{T} = (2/\pi)^2 \approx 0.41$ at $f = 1/(2T)$ and reaching the first null at $f = 1/T$. The effective detection bandwidth is therefore $\Delta f \sim 1/(2T)$. Laser phase noise cancels in the differential measurement (common-mode rejection), analogous to TDI in LISA.

\emph{The gravitational wave signal enters through Mechanism~C (light propagation in curved spacetime), not Mechanism~A (internal atomic coupling).} The no-go result of Sec.~\ref{sec:coupling} does not apply to atom interferometric detectors. The atom serves as a freely falling test mass and quantum phase reference, not as a tidal sensor.

\subsection{The mid-band frequency gap}
\label{sec:midband}

Atom interferometric detectors target $\sim$0.01--10~Hz, lying between the LISA band ($\lesssim 0.1$~Hz) and the LIGO band ($\gtrsim 10$~Hz). This gap falls outside the reach of both operating and approved instruments: LISA's sensitivity degrades above $\sim$0.1~Hz due to the arm-length transfer function, while ground-based detectors are limited below $\sim$10~Hz by seismic and Newtonian (gravity gradient) noise. A range of concepts has been proposed to target this band directly~\cite{Mandel2018}; the atom-interferometric approach is the focus here.

Atom interferometers can access this band because the atoms are in \emph{free fall} during the interrogation period, with no mechanical connection to the ground. They are immune to suspension thermal noise and partially immune to seismic noise (though gravity gradient noise remains a challenge for terrestrial implementations). The scientific targets include early inspiral of stellar-mass binary black holes (hours to days before merger in the LIGO band), intermediate-mass black hole mergers ($10^2$--$10^4\,M_\odot$), and multiband gravitational wave astronomy.

\subsection{Sensitivity and current experiments}
\label{sec:ai_sensitivity}

The strain noise PSD at the atom shot noise limit is
\begin{equation}
\sqrt{S_h(f)} = \frac{1}{n\,k_{\mathrm{laser}}\,L\,\sqrt{N_{\mathrm{atoms}}\cdot R}} \times \frac{1}{|\mathcal{T}(f)|},
\label{eq:Sh_AI}
\end{equation}
where $N_{\mathrm{atoms}}$ is the number of atoms per shot, $R$ is the repetition rate, and $\mathcal{T}(f)$ is the transfer function with bandwidth $\Delta f \sim 1/(2T)$.

MAGIS-100~\cite{MAGIS100} (Fermilab) is a 100~m pathfinder with $T = 1.4$~s; laser laboratory construction was completed in late 2025, with installation on track for 2027 and commissioning in 2028. The proposed MAGIS-km extension ($L = 1$~km, $T \approx 4.5$~s, $n \sim 10^3$, $N \sim 10^8$) achieves a broadband atom shot noise floor of $\sqrt{S_h} \sim 10^{-17}/\sqrt{\mathrm{Hz}}$ at 0.1~Hz from Eq.~(\ref{eq:Sh_AI}). Reaching the target sensitivity of $\sim 10^{-21}/\sqrt{\mathrm{Hz}}$ requires techniques beyond broadband single-loop operation: resonant-mode interrogation with $m \sim 10^2$--$10^3$ sequential loops at a fixed frequency provides an additional $\sqrt{m}$ enhancement; large momentum transfer orders exceeding $n = 10^4$ are under active development; and satellite-based configurations with $L \sim 10^4$~km (e.g., the proposed AEDGE mission) exploit the linear scaling of Eq.~(\ref{eq:Sh_AI}) with baseline. AION~\cite{AION2020} follows a staged program from AION-10 (10~m) to AION-km. ZAIGA~\cite{ZAIGA2020} is a 300~m facility in China. MIGA, a 150~m horizontal instrument using atom interferometers interrogated by a resonant optical cavity, is under construction at the LSBB underground laboratory in France~\cite{Canuel2018}.

\subsection{Gravity gradient noise: the dominant terrestrial challenge}
\label{sec:GGN}

The primary noise source for terrestrial atom interferometric detectors at low frequencies is Newtonian (gravity gradient) noise (GGN) from seismic Rayleigh waves, which directly perturb the local gravitational field experienced by the atoms.

Mitchell, Kovachy, Hahn, Adamson, and Chattopadhyay~\cite{Mitchell2022} performed a comprehensive environmental characterization of the Fermilab MINOS shaft for MAGIS-100. Their analysis establishes several critical results. At the surface, the GGN strain equivalent for a 100~m baseline is $h_{\mathrm{GGN}} \sim 10^{-17}/\sqrt{\mathrm{Hz}}$ at 1~Hz. At the 100~m depth of the MINOS shaft, the seismic spectral density lies between the Peterson New Low Noise Model (NLNM) and New High Noise Model (NHNM), with a knee frequency of $\sim$0.48~Hz.

Crucially, the GGN can be distinguished from a gravitational wave signal using the ``string-of-pearls'' technique: multiple atom interferometers distributed along the vertical baseline. Because Rayleigh-wave-induced GGN decays exponentially with depth (with scale height $\sim \lambda_{\mathrm{Rayleigh}}/2\pi$), while the gravitational wave signal varies linearly with baseline, fitting the depth-dependent phase response can suppress GGN by factors of $10^{-2}$--$10^{-3}$ with 15--20 atom clouds and realistic 1\% phase noise per interferometer~\cite{Mitchell2022}. At kilometer depth, the GGN coupling is suppressed by $\sim 10^4$ relative to the surface~\cite{Mitchell2022}.

This mitigation strategy is essential for achieving the projected MAGIS-km sensitivity and has been validated through detailed numerical simulations with measured Fermilab seismic data~\cite{Mitchell2022}.

\subsection{Quantum enhancement through spin squeezing}
\label{sec:spin_squeezing}

The atom shot noise limit in Eq.~(\ref{eq:Sh_AI}) can be surpassed through entanglement. Spin-squeezed atomic ensembles~\cite{Hosten2016,Giovannetti2004} achieve phase sensitivity $\delta\Phi = 1/N^\xi$ with $\xi > 1/2$ (standard quantum limit), approaching the Heisenberg limit $\delta\Phi = 1/N$ for $\xi \to 1$. With $N = 10^8$ atoms and 20~dB of atomic squeezing (demonstrated in laboratory settings~\cite{Hosten2016}), the strain sensitivity improves by a factor $\sim$10 beyond the standard quantum limit---directly analogous to the squeezed vacuum enhancement in optical interferometers.

To quantify the impact for MAGIS-km: the broadband atom shot noise limited sensitivity is $\sqrt{S_h} \sim 10^{-17}/\sqrt{\mathrm{Hz}}$ at 0.1~Hz with $N = 10^8$ atoms and $n = 10^3$ LMT order [Eq.~(\ref{eq:Sh_AI})]. With 20~dB of spin squeezing, this improves to $\sqrt{S_h} \sim 10^{-18}/\sqrt{\mathrm{Hz}}$. Combined with resonant-mode operation ($m \sim 10^2$--$10^3$ sequential loops at fixed frequency), the total sensitivity can reach $\sim 10^{-21}/\sqrt{\mathrm{Hz}}$, providing sensitivity to the early inspiral of stellar-mass binary black holes at cosmological distances. Spin squeezing at a fixed level of $\xi_{\mathrm{dB}}$~dB provides a fixed noise reduction factor $10^{\xi_{\mathrm{dB}}/20}$ independent of $N$, equivalent to multiplying the effective atom number by $10^{\xi_{\mathrm{dB}}/10}$. For 20~dB, this corresponds to an effective $N_{\mathrm{eff}} = 100N$---a substantial but bounded gain. However, achieving 20~dB squeezing with $10^8$ atoms in a free-fall configuration remains a major experimental challenge---current demonstrations use trapped ensembles with $10^5$--$10^6$ atoms. The path from laboratory spin squeezing to operational gravitational wave detector enhancement requires overcoming decoherence during the $\sim$1--5~s interrogation times, maintaining entanglement during large momentum transfer pulse sequences, and operating in the free-fall environment of a vertical vacuum system.

The noise architecture framework of Sec.~\ref{sec:universal} illuminates why spin squeezing is so potent for atom interferometers \emph{at design sensitivity}. If all technical noise sources (gravity gradient noise, laser frequency noise, vibrations) are suppressed below the atom shot noise floor, the quantum-accessible noise fraction approaches $\beta \to 1$: essentially \emph{all} of the residual noise is quantum projection noise. In this regime, $\mathcal{E}_{\mathrm{max}} = 1/\sqrt{1-\beta} \to \infty$---there is no architectural ceiling on quantum enhancement, in stark contrast to LISA ($\beta \sim 0.09$, $\mathcal{E}_{\mathrm{max}} \approx 1.05$). We note that LISA's $\beta$ is derived from established hardware specifications and Pathfinder flight data, while the atom interferometer $\beta \to 1$ is a projection contingent on achieving design sensitivity in detectors that have not yet operated; during commissioning, technical noises will dominate and $\beta$ will be well below unity. Nevertheless, the fundamental noise architecture is qualitatively different: atom interferometers are \emph{designed} to be quantum-noise-limited, placing them alongside ground-based optical interferometers in a regime where quantum enhancement technology has transformative potential once technical noise sources are controlled.

\subsection{Role of quantum mechanics in the detection scheme}
\label{sec:ai_quantum}

Although the gravitational wave signal enters through the classical propagation mechanism, atom interferometric detectors are genuinely quantum in five essential respects: (i) the atom must be in a quantum superposition of spatially separated states---a classical particle cannot be in two places simultaneously; (ii) the fundamental noise floor is set by quantum projection noise $\delta\Phi = 1/\sqrt{N}$; (iii) entanglement (spin squeezing) can surpass the standard quantum limit; (iv) LMT techniques exploit coherent quantum evolution through sequential photon interactions; (v) the quantum measurement scheme enables freely falling atoms as test masses, providing natural immunity to suspension thermal noise.

\section{Ground-Based Detectors}
\label{sec:ground}

The contrast with ground-based detectors is striking. For LIGO, Virgo, and KAGRA, the high-frequency sensitivity is dominated by photon shot noise~\cite{Caves1981,Braginsky1980}, with $\beta_{\mathrm{shot}} \sim 0.85$--$0.95$ depending on frequency and configuration, as detailed below. The noise architecture is fundamentally different from LISA.

The frequency dependence of the enhancement follows directly from the spectral composition of the ground-based noise budget, and it is worth making the mechanism explicit. At high frequencies ($f \gtrsim 200$~Hz) the budget is dominated by photon shot noise, which is genuinely quantum-limited~\cite{Corbitt2006}; the quantum-accessible fraction is therefore large ($\beta_{\mathrm{shot}} \sim 0.85$--$0.95$) and squeezed vacuum, which reduces shot noise directly, yields the full enhancement $\mathcal{E} = 1.6$--$2.1$. The enhancement weakens toward lower frequencies as classical and radiation-pressure noise take over. Below $\sim$100~Hz the classical noise---seismic, Newtonian (gravity-gradient), and suspension and coating thermal noise---exceeds the quantum noise, reaching about a factor of two larger in amplitude by $\sim$100~Hz in O4~\cite{Jia2024}, so the quantum-accessible fraction available to a fixed squeezing strategy falls to $\beta \sim 0.2$. Below $\sim$50~Hz quantum radiation pressure noise, which occupies the conjugate quadrature, overtakes shot noise, so fixed-angle squeezing that reduces shot noise amplifies it~\cite{McCuller2020}; in O4 the quantum noise was measured to exceed the standard quantum limit between 35 and 75~Hz~\cite{Jia2024}. The classical contributions are not addressable by squeezing at all. We emphasize that this frequency dependence is a property of the noise budget, not of the detector's response to gravitational waves: the strain transfer function appears identically in the noise and the signal referred to strain, and cancels in the enhancement ratio $\mathcal{E}$. It is the changing balance of quantum and classical noise with frequency---not the detector response---that governs where quantum enhancement is effective.

The effective squeezing, accounting for optical losses $\mathcal{L}$, modifies the squeezed quadrature variance as
\begin{equation}
V_{\mathrm{eff}} = \frac{1-\mathcal{L}}{\xi_{\mathrm{raw}}} + \mathcal{L},
\label{eq:V_eff}
\end{equation}
giving strain enhancement $\mathcal{E} = 1/\sqrt{V_{\mathrm{eff}}}$ in the shot-noise-limited regime. Table~\ref{tab:ground} evaluates $\beta_{\mathrm{shot}}$ in this regime; extending the enhancement across all frequencies, where the shot--radiation-pressure composition sets the required strategy, calls for the frequency-dependent treatment of Zhang and Miao~\cite{ZhangMiao2025}.

It is useful to give the intuitive reason that squeezing works at all, and why its benefit is tied to the noise architecture. The light entering the interferometer's dark port carries irreducible vacuum fluctuations, distributed between two conjugate quadratures (amplitude and phase) bound by an uncertainty relation. Squeezing does not add signal or reduce the total quantum uncertainty; it redistributes that uncertainty, lowering the fluctuations in the phase quadrature that carries the shot-noise-limited signal at the cost of raising them in the conjugate quadrature. Because squeezing acts only on the quantum (shot) noise, the improvement it produces is bounded by the fraction of the total noise that is quantum in origin---precisely the quantity $\beta$. The quantity being adjusted is the orientation of the squeezed quadrature relative to the interferometer readout---aligned to the phase quadrature that carries the signal, and, for broadband operation, rotated with frequency to avoid degrading the radiation-pressure-dominated band; the full experimental realization is detailed in Refs.~\cite{Tse2019,Acernese2019,Jia2024}. The physical quantity adjusted and the method of implementation are identical for ground-based detectors and LISA; what differs is the noise environment on which squeezing acts. This is why the same technology is transformative in one detector and marginal in another: in the ground-based high-frequency band $\beta_{\mathrm{shot}} \sim 0.9$, so reducing shot noise reduces almost all of the noise; for LISA $\beta \approx 0.09$, so reducing shot noise leaves the total nearly unchanged. The limitation for LISA is therefore not in the squeezing technique or its implementation, both identical to the ground-based case, but in the detector's noise architecture: there is little quantum noise for squeezing to act upon. In practice the achievable gain is further limited by optical losses, which readmit unsqueezed vacuum and are captured by the effective variance $V_{\mathrm{eff}}$ of Eq.~(\ref{eq:V_eff}); this is why realistic detectors realize of order 5.8~dB of effective squeezing rather than the larger levels generated at the source. The principal practical challenges are of this kind: minimizing optical losses along the injection and readout path, stabilizing the squeezed-quadrature angle against phase fluctuations (an error in the angle mixes the anti-squeezed quadrature back into the readout), and, for broadband operation, controlling the length and losses of the filter cavity that rotates the angle with frequency~\cite{McCuller2020,Jia2024}.

\begin{table}
\caption{\label{tab:ground}Squeezed vacuum enhancement for ground-based detectors. The enhancement $\mathcal{E}$ is computed from Eq.~(\ref{eq:enhancement}) using the representative shot noise fraction $\beta_{\mathrm{shot}}$ in the shot-noise-dominated regime ($f \gtrsim 200$~Hz)~\cite{Corbitt2006}. The squeezing levels are the effective values, i.e.\ after optical losses: the value measured at LIGO Livingston in O4 (5.8~dB, a factor of 1.9 in noise amplitude)~\cite{Ganapathy2023} and, for the third-generation detectors, the $10$~dB design target reduced by a representative $10\%$ optical loss (5\% injection and 5\% readout) to $7.2$~dB. With effective squeezed variance $\xi_{\mathrm{eff}} = 10^{-\mathrm{dB}_{\mathrm{eff}}/10}$, the enhancement is $\mathcal{E} = 1/\sqrt{1-\beta_{\mathrm{shot}}(1-\xi_{\mathrm{eff}})}$; at the highest frequencies where $\beta_{\mathrm{shot}} \to 1$, it approaches its ceiling $1/\sqrt{\xi_{\mathrm{eff}}}$ ($= 1.9$ and $2.3$ for the two levels).}
\begin{tabular}{@{}lcccc@{}}
Detector & dB$_{\mathrm{eff}}$ & $\beta_{\mathrm{shot}}$ & $\mathcal{E}$ & Volume gain \\
\hline
LIGO O4 & 5.8 & 0.85 & 1.6 & $4.4\times$ \\
ET-HF ($f \gtrsim 30$~Hz) & 7.2 & 0.95 & 2.1 & $9.0\times$ \\
ET-LF ($f \sim 3$--$10$~Hz) & --- & 0.1--0.2 & 1.05--1.12 & $1.2$--$1.4\times$ \\
Cosmic Explorer & 7.2 & 0.95 & 2.1 & $9.0\times$ \\
\end{tabular}

\end{table}

The physical root of this difference is the received optical power. Ground-based detectors operate at high power ($\sim$750~kW in Advanced LIGO, $\sim$3~MW in ET-HF), making quantum noise the dominant limitation at high frequencies; LISA receives only $\sim$700~pW after the $\sim 10^{10}$ beam-divergence loss over 2.5~Gm, so shot noise is a minor OMS contributor. This power difference, not the squeezing technology, is what places ground-based detectors and LISA in opposite noise-architecture regimes.

\emph{The Einstein Telescope xylophone design.} The Einstein Telescope~\cite{ET2010,Hild2010} provides a particularly instructive application of the noise architecture framework. Rather than building a single broadband instrument, ET adopts a xylophone configuration: two co-located interferometers optimized for complementary frequency bands~\cite{Hild2010}. ET-HF (room temperature, 3~MW circulating power, $f = 30$~Hz--10~kHz) operates in the high-$\beta$ regime where quantum noise dominates; it is an upgraded second-generation interferometer with 10~km arms and a 10~dB squeezing target. ET-LF (cryogenic at 10--20~K, silicon test masses, only 18~kW circulating power, $f = 2$--30~Hz) operates in a fundamentally different noise regime. At frequencies of 3--10~Hz where ET-LF achieves peak sensitivity, the noise budget is dominated by Newtonian noise (gravity gradient fluctuations from seismic density waves) and residual suspension thermal noise---both classical in origin and not addressable through quantum enhancement. The low circulating power was chosen precisely to suppress quantum radiation pressure noise at these frequencies, but it also renders shot noise subdominant. The quantum-accessible noise fraction for ET-LF is therefore small, $\beta_{\mathrm{quantum}} \sim 0.1$--$0.2$, giving $\mathcal{E}_{\mathrm{max}} \sim 1.05$--$1.12$---comparable to LISA rather than to ET-HF [see Ref.~\cite{Korobko2025} for a detailed analysis of quantum noise sources in ET].

This split is, in effect, an engineering response to the frequency dependence of $\beta(f)$: building two instruments each optimized for its noise regime is more effective than broadband quantum enhancement of a single detector spanning four decades in frequency. For ET-HF the 7.2~dB effective squeezing yields $\mathcal{E} \approx 2.1$ and a $\sim\!9\times$ survey-volume gain, whereas for ET-LF cryogenic thermal-noise suppression, underground siting, and Newtonian-noise subtraction---not quantum enhancement---are the enabling technologies.

\emph{Note on mechanical squeezing.} The feedback-stabilized parametric squeezing demonstrated by Poot, Fong, and Tang~\cite{Poot2015}, achieving 15.1~dB of noise reduction in an opto-electromechanical resonator, operates in a fundamentally different regime. This is \emph{classical} squeezing of thermal mechanical motion, not quantum squeezing of optical vacuum fluctuations. The technique reduces the thermal noise variance in one quadrature of a mechanical mode at room temperature (thermal occupation $\bar{n} \gg 1$), whereas squeezed vacuum injection in gravitational wave detectors reduces the zero-point quantum fluctuations entering the interferometer readout port. The 15.1~dB classical figure is therefore not comparable to the 6--9~dB quantum squeezing relevant to LIGO and future detectors. Mechanical squeezing of this type would be relevant only to hypothetical resonant-mass detectors operating near the quantum ground state, a regime not yet achieved for macroscopic mechanical systems at gravitational wave frequencies.

\section{Discussion and Conclusions}
\label{sec:discussion}

We have derived how gravitational waves couple to quantum systems through three distinct mechanisms, with transducer gains differing by up to $10^{35}$, and shown that this coupling---not the sensor's quantum performance---determines whether a given quantum technology can contribute. The analysis spans the observable spectrum from millihertz to kilohertz, with the mid-band gap and the nanohertz regime. This hierarchy produces the following results.

\emph{First}, direct quantum detection through internal atomic coupling is limited by a transducer gain $G_A = 2.4\times 10^{-20}$ at millihertz frequencies---a value derived from the complete three-dimensional matrix element, not from dimensional analysis. The exact vanishing of the first-order energy shift for $J = 0$ clock states (Wigner--Eckart theorem) forces the leading signal to second order in $h_0$, compounding the $\sim\!10^{35}$ gap relative to light-propagation interferometry. Even polar molecules, which evade the selection rule and recover 70 orders of magnitude over atomic clocks, remain 31 orders short.

\emph{Second}, LISA's noise architecture leaves limited room for quantum enhancement. Time-Delay Interferometry suppresses laser frequency noise to $\beta_{\mathrm{laser}} = 5\times 10^{-5}$ of the OMS budget, so that atomic clock stabilization of the laser yields $\mathcal{E} = 1.00003$. The optical metrology system leaves only a small fraction of the noise power---photon shot noise, bounded to $\lesssim 15\%$---accessible to identified quantum technologies. The combined quantum enhancement saturates at $\mathcal{E} \approx 1.04$ even with arbitrarily perfect quantum sensors.

\emph{Third}, ground-based detectors illustrate the opposite regime of the same coupling physics. These detectors also exploit Mechanism~C, but with high circulating power ($\sim$750~kW in Advanced LIGO), quantum noise constitutes the dominant limitation at high frequencies ($\beta_{\mathrm{shot}} \sim 0.85$--$0.95$). Squeezed vacuum provides $\mathcal{E} = 1.6$--$2.1$ and survey volume gains of $4$--$9\times$~\cite{Ganapathy2023,Jia2024}; frequency-dependent squeezing extends this enhancement to lower frequencies where radiation pressure noise must also be addressed. The dramatic contrast with LISA---the same coupling mechanism and the same quantum technology yielding transformative gains in one detector and marginal gains in another---demonstrates that the noise architecture controls the ceiling on enhancement, while the quantum technology determines the gain achieved within it. The Einstein Telescope xylophone design~\cite{Hild2010} embodies this principle: its high-frequency interferometer (ET-HF, $\beta \sim 0.95$) benefits from 10~dB squeezing, while its cryogenic low-frequency interferometer (ET-LF, $\beta \sim 0.1$--$0.2$) gains almost nothing from quantum enhancement---both within the same observatory.

\emph{Fourth}, atom interferometric detectors provide the clearest illustration of why the coupling mechanism matters. They succeed not because they use better quantum sensors, but because they exploit Mechanism~C (light propagation in curved spacetime) through a detection scheme in which quantum superposition, entanglement, and coherent manipulation are essential. They target the 0.01--10~Hz mid-band gap between LISA and LIGO, not covered by operating or approved instruments, and enable multiband gravitational wave astronomy. The environmental challenges, particularly gravity gradient noise, can be addressed through the vertical gradiometer array introduced by Chaibi et al.~\cite{Chaibi2016} and characterized for MAGIS-100 by Mitchell et al.~\cite{Mitchell2022} with projected suppression factors of $10^{-2}$--$10^{-3}$~\cite{Mitchell2022}. Quantum enhancement through spin squeezing provides an additional $\sim$10$\times$ improvement, analogous to squeezed vacuum in optical detectors.

\emph{The nanohertz band.} Pulsar timing arrays (PTAs) detect gravitational waves at $f \sim 10^{-9}$--$10^{-7}$~Hz by monitoring radio-pulse arrival times from millisecond pulsars across the Galaxy; the NANOGrav, EPTA, PPTA, and IPTA collaborations have reported evidence for a stochastic background in this band~\cite{NANOGrav2023}. PTAs exploit Mechanism~C: the wave perturbs the pulse propagation over the kiloparsec baseline between pulsar and Earth, producing a correlated modulation of arrival times (the Hellings--Downs curve). The pulsar is the test mass and the radio wave is the light---identical to LISA's Mechanism~C, with the baseline extended from gigameters to kiloparsecs.

The noise architecture, however, is qualitatively distinct. The dominant sources---intrinsic pulsar spin noise, dispersion measure variations from the interstellar medium, solar wind fluctuations, and the finite number of pulsars with sub-microsecond precision---are astrophysical, not quantum-mechanical. The only quantum-accessible component is radiometer noise in the receiver backend, subdominant for well-timed pulsars, so quantum-limited amplifiers would not help. This is a third regime: $\beta_{\mathrm{quantum}} \approx 0$ not from optimized classical engineering (as in LISA) nor architectural splitting (as in ET's xylophone), but because the noise is astrophysical. PTA sensitivity improves through more stable pulsars, longer timing baselines, and better chromatic-noise modeling---not quantum enhancement.

\emph{Quantum networks of clocks.} Entangled clock networks~\cite{Komar2014} could improve Mechanism~B beyond the standard quantum limit: Heisenberg-limited entanglement scales the strain sensitivity as $N$ rather than $\sqrt{N}$. For $N = 100$ entangled clocks this gains a factor of 100 over a single pair, improving $h_{\mathrm{min}}^{\mathrm{dilation}} = 7.1\times 10^{-18}$ to $\sim 7\times 10^{-20}$---still a factor of $\sim 70$ above the LISA target. Even with 100 space-qualified clocks holding Heisenberg-limited coherence over interplanetary baselines, the coupling mechanism, not the clock performance, sets the sensitivity floor.

\emph{Other quantum matter proposals.} Several proposals couple to quantum matter beyond atomic clocks. Sab\'{i}n et al.~\cite{Sabin2014} proposed detecting gravitational waves through phonon creation in a Bose--Einstein condensate, interpreted as a dynamical Casimir process. The mechanism is quite elegant; we observe though that it couples through sound-speed modulation of the condensate---a tidal deformation of its collective modes, hence Mechanism~A---and the coupling scales with the condensate size ($\sim 10^{-4}$~m), still negligible against the gravitational wavelength, yielding strain sensitivities many orders of magnitude above relevant levels. Optomechanical proposals~\cite{Arvanitaki2013} using levitated dielectric microspheres target the ultrahigh-frequency band ($\gtrsim$~kHz), coupling through Mechanism~A at the center-of-mass level and facing the same transducer gain hierarchy, with a target range outside the present scope.

More recently, Tobar et al.~\cite{Tobar2024} have proposed detecting single gravitons using massive quantum acoustic resonators (kg-scale bars) cooled to their quantum ground state. Their starting point is the same atomic-scale coupling deficit quantified in Sec.~\ref{sec:coupling}: Weinberg's calculation~\cite{Weinberg1972} gives a spontaneous graviton emission rate $\Gamma_{\mathrm{spon}} \sim 10^{-40}$~Hz for the hydrogen $3d \to 1s$ transition---the matrix element we compute in Sec.~\ref{sec:matrix_element}. They circumvent it by moving to macroscopic scales, replacing the atomic scale $m_e a_0^2$ with $M L^2$: for a kg-scale, meter-scale bar the mass ratio alone gives $M/m_e \sim 10^{30}$ and the geometric ratio $(L/a_0)^2 \sim 10^{20}$, exceeding the atomic coupling by many tens of orders of magnitude. Even so, autonomous detection is not feasible: their proposal correlates quantum jumps in the resonator with \emph{classical} LIGO detections of the same event, with LIGO as the herald and the resonator probing whether the exchanged energy is quantized. This confirms the transducer gain hierarchy at all scales---Mechanism~A, regardless of mass and size, does not reach the strain sensitivity of Mechanism~C. The goal of such experiments is not gravitational wave astronomy but tests of quantum gravity, a distinct objective; a comprehensive review of massive quantum systems as probes of gravity is given by Bose et al.~\cite{Bose2025}.

\emph{Quantum enhancement frameworks.} For quantum noise reduction in laser interferometers, Zhang and Miao~\cite{ZhangMiao2025} have developed a comprehensive framework based on the fundamental quantum limit set by stored optical energy, unifying squeezing, variational readout, speed meters, and related techniques for ground-based detectors. Our noise architecture analysis ($\mathcal{E}_{\mathrm{max}} = 1/\sqrt{1-\beta}$) addresses a different question: once the coupling mechanism is viable (Mechanism~C), what fraction of the total noise budget is quantum-accessible? This applies across all detector architectures and frequency bands, including space-based and atom interferometric detectors where the answer differs dramatically from ground-based interferometers.

The lesson throughout is the same: the coupling mechanism determines the outcome. Proposals that couple through internal atomic structure (Mechanism~A) face a $10^{35}$ transducer gain deficit set by the coupling, not by sensor performance. Proposals that couple through center-of-mass motion (Mechanism~B) reach strain sensitivities orders of magnitude above astrophysical requirements. Proposals that couple through light propagation over macroscopic baselines (Mechanism~C) achieve the transducer gain needed for gravitational wave astronomy---and for these, the quantum improvement available is set by the detector's noise architecture at the frequency of interest. The first question is not ``how good is the quantum sensor?'' but ``how does the gravitational wave couple to the measurement?'' Only after the coupling mechanism establishes viability does the sensor's performance become the determining factor.

\section*{Acknowledgments}
The author acknowledges the use of Anthropic Claude as a writing assistant in the preparation of this manuscript.

\section*{Data availability statement}
This is a theoretical study; no new experimental or observational data were created or analysed. The analytic derivations underlying the results and the code that reproduces all numerical values and the figure are openly available in Zenodo at \url{https://doi.org/10.5281/zenodo.20749698}.

\section*{Funding}
This research received no external funding.

\section*{Conflicts of interest}
The author declares no conflicts of interest.

\end{document}